\documentclass[12pt,article]{JHEP3}
\usepackage{amsmath,amssymb,euscript,array,mathrsfs,epsfig}
\newcommand{\startappendix}{
\setcounter{section}{0}
\renewcommand{\thesection}{\Alph{section}}}
\newcommand{\Appendix}[1]{
\refstepcounter{section}
\begin{flushleft}
{\large\bf Appendix \thesection: #1}
\end{flushleft}}
\def\ben
{\begin{equation}}
\def\een{\end{equation}}

\def\be{\begin{equation}}
\def\ee{\end{equation}}
\def\ba{\begin{array}}
\def\ea{\end{array}}

\def\dalemb#1#2{{\vbox{\hrule height .#2pt
        \hbox{\vrule width.#2pt height#1pt \kern#1pt
                \vrule width.#2pt}
        \hrule height.#2pt}}}

\newcommand{\bea}{\begin{eqnarray}}
\newcommand{\eea}{\end{eqnarray}}

\newcommand{\Tr}{{\rm Tr} }
\newcommand{\sh}[1]{#1\hskip-7pt /}

\def\vep{{\varepsilon}}

\title{Spinning flavour branes and fermion pairing instabilities}
\author{$\quad$ S. Prem Kumar\\\\
{\it 
Department of Physics, \\
Swansea University, \\ 
Singleton Park, Swansea\\
SA2 8PP, U.K.  
}\\
E-mail: \email{s.p.kumar@swansea.ac.uk}
} 
\abstract{We consider probe Dp-branes, $p=3,5,7$, in global $AdS_5\times S^5$,
  rotating along an internal direction in the $S^5$. These are dual to 
  strongly interacting ${\cal N}=4$ SYM on $S^3$ with massless 
  fundamental flavours, in
  the presence of an R-symmetry chemical potential for flavour fermions. 
  For massless,  
  ``AdS-filling" Dp-brane embeddings at zero temperature, 
  we find an infinite set of 
  threshold values of the  chemical potential at which instabilities
  are triggered. The
  onset of instability is always preceded by metastability of the zero
  density state. From the onset values of the chemical potential, we
  infer that  unstable directions favour a homogeneous 
condensate of a bilinear made from fermion harmonics, or  
Cooper pairing. 
We confirm this picture both numerically and 
  analytically. The linearized analysis showing the appearance of 
 instabilities involves a charged scalar in global AdS space coupled
  to a (large) constant background gauge potential. The resulting
  frequency space correlator of the fermion bilinear at strong
  coupling displays poles in the upper half plane.  In contrast, the
  correlator at zero coupling  exhibits Pauli blocking due to occupation
  of states below the Fermi level,  but no instabilities.  The  
 end-point of the strong coupling instability is not visible in our setup. 
}
\begin{document}
\section{Introduction and summary}
The behaviour of strongly coupled systems under extreme  
densities and temperatures is important in a variety of
physical situations. Gauge/gravity dualities
\cite{maldacena, magoo, witten1} present us with an approach towards
tackling  the physics of such situations, both for pure gauge theories
\cite{witten1, gubser} 
and gauge theories with quark-like matter 
\cite{karchkatz, mateos1, mateos2, ss, Aharony:2006da, horigome}. 
Holographic duals of strongly interacting fermionic systems at 
finite density are of considerable interest in the context of
superconducting/superfluid systems \cite{sean}. Such dual descriptions
can also provide the necessary window into the workings of strongly
interacting quark matter. This is particularly relevant for QCD where
phase transitions are conjectured to occur precisely at densities
where the coupling is strong, rendering perturbative methods
ineffective, and where lattice approaches need to surmount the
technical obstacle posed by the sign problem.
 
In this note we explore properties of one such setup involving ``probe'' 
Dp-branes
$(p=3,5,7)$ in {\em global} $AdS_5\times S^5$, with the branes chosen to
spin along a compact internal direction in the $S^5$. The dual
conformal field theories live on spheres. Imparting a rotation to 
the Dp-branes in the internal directions allows for a chemical
potential (conjugate to a $U(1)$ R-charge) for
massless fermion flavour fields transforming in the fundamental representation
of the gauge group in the boundary field theory.

One of our original 
motivations for examining these systems in global AdS spacetime was to
make contact with  recent studies of the phase diagram of $SU(N)$ gauge
theories in finite volume with fermion flavours
in the fundamental representation \cite{Hands:2010zp,
  Hands:2010vw}. However, aspects of our setup and the resulting
physical picture are possibly also of relevance to ``top-down'' approaches
towards holographic superconductor/superfluid models 
involving probe branes (e.g. 
\cite{O'Bannon:2008bz, Ammon:2008fc,Ammon:2009fe, Ammon:2010pg, 
Myers:2008me,Basu:2008bh}).

The works of \cite{Hands:2010zp,
  Hands:2010vw} found, using a weak coupling analytical approach at large-$N$, 
and small volume lattice simulations of QCD, an
exotic phase structure of $SU(N)$ gauge theories 
as a function of quark/fermion number
chemical potential. It was shown that at large-$N$ with a
large number of flavours, the systems exhibit an infinite
sequence of third-order Gross-Witten transitions, across which, fermion number
increases in discrete steps. A jump in fermion number occurs when
the chemical potential crosses the energy of an elementary fermion
harmonic on the sphere, and the corresponding 
energy level gets occupied. This jump
is accompanied by a spike in the Polyakov loop order parameter and a
pair of third-order transitions \footnote{Lattice
studies of QCD at finite volume \cite{Hands:2010vw}, 
showed the same qualitative jumps in quark number
correlated with spikes in the Polyakov loop. However, these studies
also found certain qualitative differences including 
the appearance of metastable configurations.}.

It is natural to ask whether the physics of fermions seen above, 
can be observed in a 
setting that has a strong coupling description 
within a gravity dual framework. The probe brane configurations we
consider present us with such a setup. However, while
\cite{Hands:2010zp, Hands:2010vw} treated theories with $N_f$ and
$N$ comparable, we will only 
explore the theory in the 't Hooft large-$N$ limit with $N_f/N \to 0$.

The main result of our study is that the strongly coupled system in
finite volume does not see phenomena associated to the energy
levels of individual, elementary fermion states. Instead, the system is
unstable to the condensation of composite scalars made from bilinears of
fermion harmonics on the sphere. The transitions associated to this
occur when the chemical potential times the R-charge approaches 
twice the energy of a
fermion harmonic, eventually triggering an instability towards fermion
bilinear condensation. The strong coupling analysis shows that a large
enough R-symmetry chemical potential acts as a negative mass squared for the
fermion bilinears. As expected, the instabilities persist in the
infinite volume limit, and potentially imply a phase with
spontaneously broken $U(1)$ R-symmetry in the presence of a chemical
potential. A complete description of such a symmetry breaking ground
state would provide a holographic model of a
superfluid/superconductor. 
However, within our probe prane ansatz, 
the instabilities appear to be of a runaway nature and
the end-point solution is not visible in the setup. Investigations of
spinning flavour branes in the Poincare' patch of $AdS_5$ were first
carried out in \cite{evans}, while a different way of
inducing R-symmetry breaking in the spinning brane setup, via a 
probe brane magnetic field, was investigated in \cite{O'Bannon:2008bz}.

The spinning probe Dp-branes considered here, 
describe strongly coupled ${\cal N}=4$ supersymmetric $SU(N)$ Yang-Mills
theory on the three-sphere, coupled to flavour multiplets in a
way that preserves eight supersymmetries (${\cal N}=2$ in four
dimensions). When the flavours are massless, the theories possess a
global R-symmetry.
The angular velocity of the probe D-branes along a
compact internal direction acts as a chemical potential for this
R-symmetry under which fermions in the flavour multiplets are charged.
In the strict limit $N\to \infty$ and $N_f$ fixed, which is the probe
  limit for the Dp-branes, the R-symmetry is
  non-anomalous and is a conserved quantum number. Furthermore, since
  the flavours do not ``backreact'' on the pure gauge dynamics in this
  limit, we may consistently choose all the R-charge to be concentrated in the
  flavour sector alone. The ``leaking'' of the charge to adjoint matter
  fields of ${\cal N}=4$ SYM is suppressed in the large-$N$ limit. In
  the gravity picture this means that we can have the probes spinning
  along the internal directions, whilst the supergravity background is
 not boosted along the same internal directions.

In addition, in the probe limit,  
the corresponding field theories remain conformally invariant. The
D7-brane probe yields a four dimensional ${\cal N}=2$ SUSY gauge
theory, whilst the D5- and D3-brane probes give rise to 2+1 and 1+1
dimensional ``defect'' CFTs respectively
\cite{DeWolfe:2001pq,Constable:2002xt}. 
In the global AdS
description, the defect CFT's are localized on an equatorial $S^2$ and
$S^1$ of the boundary three-sphere, and the degrees of freedom
localized on the defects interact with the ${\cal N}=4$ theory on $S^3$.

\begin{figure}[h]
\begin{center}
\epsfig{file = 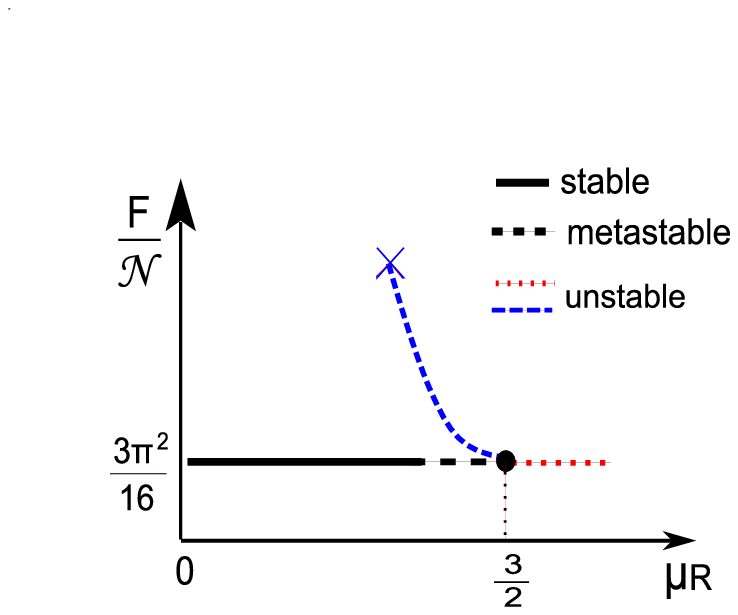, width =2.8in}
\hspace{0.5in}
\epsfig{file = 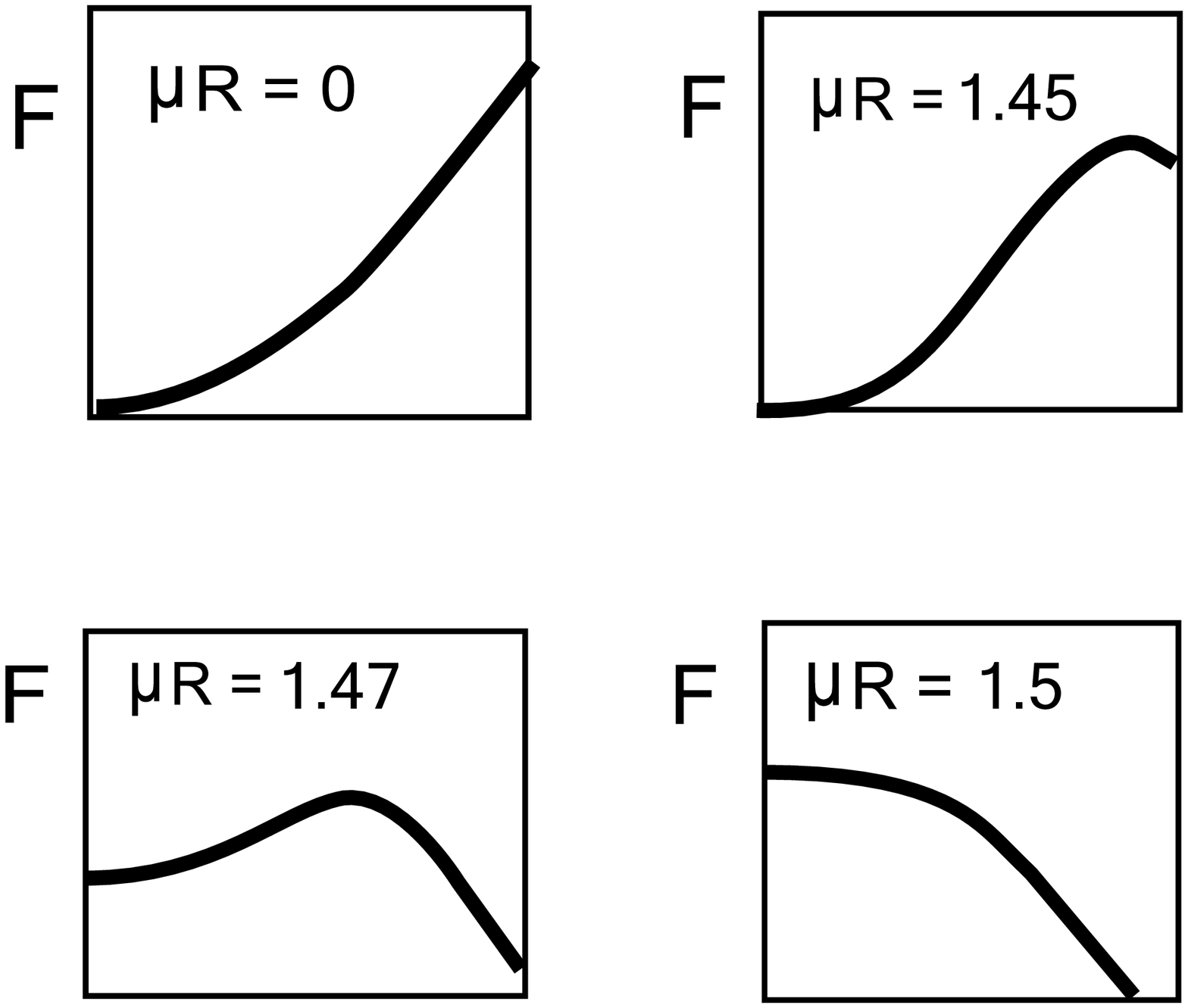, width =2.5in}
\end{center}
\caption{ 
\small{
{\underline{Left}}: The free energy as a function of chemical
  potential $\mu$ for the D7 system. The zero density state (solid
  black) is locally
  stable as $\mu$ is increased from zero. A new unstable finite
  density saddle point (dashed blue) with chiral condensate appears
  when $\mu$ approaches $\frac{3}{2}R^{-1}$. 
The two merge at $\mu R=\frac{3}{2}$ where an
  instability develops. {\underline{Right}}: The situation is schematically depicted
  in terms of an effective potential (see also Fig.8). 
}
}
\label{unst} 
\end{figure}
The strong coupling analysis of the probe Dp-brane action 
reveals that the massless, zero density embedding 
encounters an infinite sequence of instabilities  with
increasing chemical potential $\mu$, for the $U(1)$ R-symmetry. In particular
the instabilities favour the condensation of fermion bilinears with R-charge $Q_R=+2$, made
up of harmonics of elementary fermions on the sphere. 
For the D7-branes we find instabilities at 
\be
\mu\,Q_R\, = \,(2\ell +1)\,R^{-1}\,,\qquad \ell =1,2,3,\ldots
\ee
while for the defect CFT's on $S^2$ and $S^1$, the respective onset
values are 
\be
\mu\,Q_R\,=\,2\,\ell\,R^{-1} \qquad{\rm and}\qquad 
\mu\,Q_R\, =\, (2\ell-1)\,R^{-1}\,,\qquad
\ell=1,2,3,\ldots. 
\ee
The R-charge of the bilinears is $Q_R=2$, and 
in each case the critical value of $\mu\,Q_R$ is twice the energy of
an elementary fermion harmonic (on $S^3$, $S^2$ and $S^1$). 

The solutions to the non-linear equations of motion of the DBI action,
which we analyze numerically,
display a somewhat intricate structure in the vicinity of each of
these instabilities. The zero density state, which starts off being
stable, actually becomes {\em metastable} before the onset of the
first instability, accompanied by the appearance of a new thermodynamically
{\em unstable} solution with finite density and fermion bilinear
condensate. This latter saddle point eventually merges with the zero
density solution at the critical value of the chemical
potential. Beyond this value the zero density solution is simply
unstable. This phenomenon is summarized in Fig. 1 and in Fig. 8.

Analytical solutions of the linearized equations of motion for the Dp-branes
show the emergence of non-trivial embeddings (at linear order)
only at the critical values of the chemical potential. 
This confirms the result of the numerical analysis which points
towards a merger of two saddle points 
(a zero density trivial solution and 
a non-trivial finite density solution) at the critical values of the
chemical potential. The linearized equations which lead to non-trivial
massless embeddings, are precisely the equations of motion of a
charged scalar in global AdS spacetime, coupled to a constant
background $U(1)$ gauge potential. For large enough, special
values of this background gauge field (the chemical potential), the
equations allow for charged fermion bilinear condensates or ``scalar
hair''.  In this
respect the systems are somewhat 
similar to holographic superconductor models
\cite{sean,hintro}.

The linearized approach also allows the
computation of the two-point function for the fermion bilinear which
displays an infinite set of discrete poles in the complex frequency
plane. At $\mu=0$, the locations
of the poles are at twice the energies of the discrete fermion
harmonics. With a non-zero chemical potential, the poles migrate and
eventually move to the upper half plane, signalling instabilities 
when onset values of the chemical potential are crossed.

While the detailed features associated to the fermion bilinear
instability follow from our anstaz for the probe D-brane action (which
only incorporates the so-called ``slipping mode'' degree of freedom),
the end-point of these instabilities does not seem to be captured by
this action. This suggests that new effects must be included to obtain
a complete description.

The paper is organized as follows. For the purpose of completeness, in
Section 2, we
review basic field theory features of the probe D7-brane system,
paying attention to the introduction of an R-charge chemical
potential for flavours alone. In Section 3, we explain weak
coupling features of the theory on a sphere with 
non-vanishing chemical potential. The probe D7-brane analysis is done
in Section 4, with numerical results for the phase diagram and free
energies of solutions.
Analytical results in the linearized
approximation and strong-coupling Green's functions are derived in
Section 5. We discuss the rotating D5- and 
D3${}^\prime$-branes, related to D3/D5 and D3/D3${}^\prime$ systems
 in Section 6, and conclude with
speculations on the end-point of the instabilities in Section
7. Various details of weak coupling calculations of the phase structure
on $S^3$, Green's functions at finite density, and technical
features of the D7-brane ansatz are  presented in an Appendix.

\section{Adding flavours to ${\cal N}=4$ SYM on $S^3\times {\mathbb R}$}
It is well known that the simplest way of incorporating flavour fields
into the AdS/CFT framework is via probe Dp-branes in the bulk type IIB
supergravity background dual to, for example, ${\cal N}=4$ SYM with
$SU(N)$ gauge group \cite{karchkatz}. The ``probe'' approximation is
valid when the the number of flavour multiplets  $N_f$ is fixed in the 't Hooft
large-$N$ limit.

  The presence of flavour fields spoils conformality of the ${\cal
  N}=4$ theory since the gauge coupling now runs and has a putative
  Landau pole in the ultraviolet (UV). In the limit $N_f/N \to 0$
  however, the loop contributions from flavour modes are suppressed
  and the beta function for the 't Hooft coupling, $\lambda=g^2_{YM}
  N$, is proportional to $\frac{N_f}{N}$ and therefore vanishes in the
  strict large $N$ limit. We may continue to view the theory as being
  conformal, provided the hypermultiplets have vanishing bare
  masses. It is natural for CFTs in four dimensions to be viewed as
  field theories formulated on a spatial three-sphere, a picture that
  follows from radial quantization on ${\mathbb R}^4$. The radius of
  the three-sphere provides a new scale, allowing for the possibility
  of non-trivial thermodynamics as a function of temperature and/or
  chemical potentials \cite{Sundborg:1999ue, Aharony:2003sx}.

There are three distinct cases to be considered, corresponding to
probe D7-, D5- and D3-branes. In all cases the resulting system
preserves 8 supersymmetries. The D7-brane probes describe the dynamics
of ${\cal N}=2$ SUSY gauge theory with $N_f$ hypermultiplets, while
the dynamics of the two other kinds of probe branes yields lower
dimensional defect conformal field theories \cite{DeWolfe:2001pq,
  Constable:2002xt}. 
We  will begin with the study of the D7-brane probes. 

\subsection{D7-brane probes: Matter content and symmetries}
The matter content and the interactions of the field theory are most
easily summarized in the language of ${\cal N}=1$ superfields on
${\mathbb R}^{3,1}$. Each ${\cal N}=2$ hypermultiplet consists of two
${\cal N}=1$ chiral superfields 
$(Q,\tilde Q)$ transforming in the $(N, \bar N)$
representation of the $SU(N)$ gauge group. The superfields in the
hypermultiplet, each consist of a squark and a quark $Q\equiv (q,\psi_q)$ and
$\tilde Q\equiv (\tilde q, \tilde \psi_{q})$. The quarks $\psi_{q}$ and 
$\tilde \psi_{q }^\dagger$ are left- and 
right-handed Weyl fermions respectively.

With $N_f$
hypermultiplets $(Q^i,\tilde Q_i)$ and using the ${\cal N}=1$ SUSY notation,
the superpotential of the theory with flavours on ${\mathbb R}^{3,1}$, 
can be written as
\be
W= \frac{1}{g^2_{YM}} \left(\sum_{i=1}^{N_f}\left(\sqrt 2 \,\tilde
    Q_i\,\Phi_3\, Q^i + m \tilde Q_i\,Q^i\right) + \sqrt 2 \,{\rm
    Tr}\,(\Phi_3\,[\Phi_1,\Phi_2])\right). 
\label{superpot}
\ee
 The six adjoint scalars of the ${\cal N}=4$ theory, 
and their superpartners are packaged into the 
three (complex)  chiral multiplets $\Phi_1, \Phi_2$ and $\Phi_3$. 
Upon coupling to the ${\cal N}=2$ hypermultiplets as in the above
superpotential,  
$\Phi_3$ naturally gets identified with 
the scalar component of the ${\cal N}=2$ vector
multiplet, while $\Phi_{1,2}$ together make up an ${\cal N}=2$
adjoint hypermultiplet.  

Although one may allow for bare masses for the hypermultiplets, we 
 will only be interested in the theory with massless
 quarks. In this case the theory has the following global symmetry group
\be
{
G_F \simeq U(1)_B\times SU(N_f)\times 
SU(2)_R \times U(1)_R\times SU(2)_{\Phi}.
} 
\ee
The multiplets $Q^i$ transform in the fundamental representation
 of the flavor symmetry $U(N_f)\simeq U(1)_B\times SU(N_f)$, 
 while $\tilde Q_i$ transform in the conjugate
 representation. 
The $SU(2)_R\times U(1)_R$ factor is the R-symmetry group of the ${\cal N}=2$
gauge theory. The $SU(2)_R$ subgroup acts on the doublets
$(q^i,\tilde q^{\,\dagger \,i})$ and leaves all other fields invariant. We
 will be interested in the $U(1)$ subgroup of the R-symmetry. 
Under a $U(1)_R$ transformation, the scalar component of
$\Phi_3$ has charge $+2$, whilst the squarks 
in the hypermultiplets are uncharged. Importantly,
the fermions $\psi_{q}^i$ and ${\tilde{\psi}_{q\,i}^\dagger}$ each carry
R-charges $+1$ and $-1$  respectively. The $U(1)_R$ rotation is
therefore an axial symmetry acting on the fermions
\footnote{We take the fermions $\psi_q^i$ and $\tilde{\psi}_{ q \, i}^\dagger$
 to be left and right handed chiral fermions respectively. This is
 consistent with the Yukawa coupling implied by the superpotential
 \eqref{superpot}.}.

The $U(1)_R$ symmetry is anomalous for a fixed number
of colours and flavours. However, ${\cal N}=2$ 
supersymmetry relates this anomaly to the beta function for the gauge
theory. Like the beta function, the $U(1)_R$ anomaly is also
parametrically suppressed in the
large $N$ limit with $N_f$ fixed. It is thus
consistent to treat the $U(1)_R$ as a symmetry of the theory in the 
'tHooft large-$N$ limit. 

\subsection{The theory on $S^3\times {\mathbb R}$}
When the ${\cal N}=2$ SUSY gauge theory described above is formulated
on a spatial three-sphere, conformal invariance requires that we 
add to the Lagrangian of the theory, a
coupling of the scalar fields to the Ricci scalar of $S^3$:
\be
{\cal L} \to  {\cal L} + \frac{1}{R^2}\;\left[
\sum_{a=1}^3 {\rm Tr} \,|\Phi_a|^2 + \sum_{i=1}^{N_f} (|q_i|^2 +
|\tilde q_i|^2)\right].
\ee
Here $R$ is the radius of the boundary three-sphere.
The conserved current associated to the $U(1)_R$ symmetry is a sum
of two parts: a contribution $J_A$ from the charged adjoint matter of
the ${\cal N}=4$ sector and another piece $J_F$ from the R-charged
fundamental fermions. 
\bea
&&j^\mu_R = J^\mu_A + J^\mu _F \\\nonumber\\\nonumber
&& J^\mu_A = i\,\Tr\left(2\, \Phi_3^\dagger {\overleftrightarrow{\nabla}}^\mu \Phi_3 + 
\psi^\dagger \bar{\sigma}^\mu \psi+
\lambda^\dagger \bar{\sigma}^\mu \lambda\right)\\\nonumber
&& J^\mu_F = - i \,\sum_{j=1}^{N_f}\left(\psi_{qj}
  ^\dagger\,\bar{\sigma}^\mu \,\psi_{q}^j  
- \,{\tilde{\psi}_{q j}}\,{\sigma}^\mu \,\tilde{\psi}_{q}^{\,j\,\dagger}
\right)\,.
\eea
Chemical potentials for global symmetries can be turned on in the
Lagrangian formulation by first imagining that the corresponding
global symmetry is gauged and then introducing a background value for
the time component of that gauge field. For the $U(1)_R$ symmetry
above, a chemical potential will shift the Lagrangian as,
\be
{\cal L} \to {\cal L} - \mu (J_ A^0 + J_F^0) -\mu^2 |\Phi_3|^2\,.
\ee
As is well known \cite{Yamada:2006rx, Yamada:2007gb,
  Hollowood:2008gp}, for $\mu > 1/R$, and in the absence of
fundamental flavours,  
the chemical potential  induces a runaway instability in the 
potential energy for the adjoint 
scalar $\Phi_3$. However, this issue will not be a problem for us as we
explain below.

As we will work in the probe limit for flavour branes, 
the number of flavour
hypermultiplets is small, $N_f \ll N$, so that the flavours 
do not ``backreact'' on the ${\cal N}=4$ gauge theory. 
We then introduce a chemical potential for
the $U(1)_R$ charge in the flavour sector
alone. Strictly speaking, this will not result in an equilibrium
configuration since the R-charge carried by the flavour degrees of
freedom will eventually leak out to the R-charged adjoint modes of the
full theory
\footnote{The spinning D-brane setup has been applied to understand
  aspects of this non-equilibrium system with {\em massive} hypermultiplets
in flat space \cite{Das:2010yw}. In this case the rotation of the
branes gives rise to a time-dependent phase for the mass.}. 
However, in the 't Hooft large $N$ limit with $N_f$
fixed, this effect is suppressed. 
\begin{figure}[h]
\begin{center}
\epsfig{file = 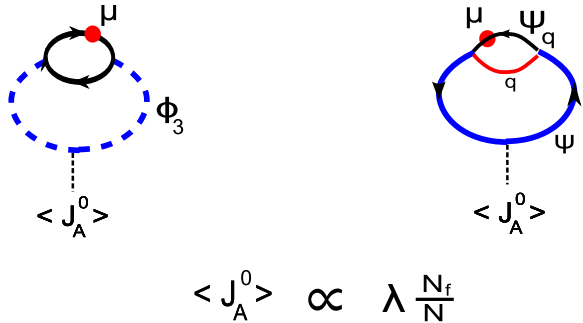, width =3in}
\end{center}
\caption{ \small{Typical graphs that contribute to a one-point function for
  $J_A^0$, the R-charge density 
in the adjoint sector, when a chemical potential $\mu$ for the flavour R-charge
  density is introduced.  $\Phi_3$ and $\psi$ are the complex adjoint
  scalar and its superpartner, which are both charged under $U(1)_R$,
  while $(q,\psi_q)$ represent flavour squarks and quarks respectively.}}
\label{conserved} 
\end{figure}

As shown in Fig. \eqref{conserved}, a chemical potential conjugate to
the flavour R-charge density $J_F^0$ also implies a non-zero one-point
function for R-charge density for adjoint matter. These
effects are suppressed by flavour loops in the 'tHooft large-$N$
limit. Therefore in this limit, it will be consistent for us to
introduce an R-charge chemical potential in the flavour sector alone,
\be
{\cal L} \to {\cal L} - \mu\, J_F^0\,.
\ee

\section{Weak coupling interlude}
\label{weak}

For the weakly coupled theory on $S^3$, the energies of all the
perturbative modes are quantized and there are no zero modes (at zero
temperature). The putative zero modes for scalars are lifted by the
conformal coupling of the scalars to the curvature of the
sphere. The eigenfunctions of the spinor Laplacion on $S^3$
have the eigenvalues and degeneracies (see e.g. \cite{sbh})
\be
\varepsilon_\ell= \left(\ell +\tfrac{1}{2}\right)R^{-1}\,,\qquad d_\ell =
2\ell(\ell+1)\,,\qquad\ell=1,2,\ldots
\label{fermion}
\ee
At large-$N$, for a non-anomalous R-symmetry, 
the chemical potential
for the $U(1)_R$ symmetry in the flavour sector is conjugate to the
chiral fermion number, which is the number of left-handed Weyl
fermions $N_L$ minus the number of right-handed fermions $N_R$. At
zero temperature, as $\mu$ is increased past each of the energy levels
\eqref{fermion}, the quantity $N_L-N_R$ will undergo a discrete jump,
\be
\Delta (N_L-N_R) = d_\ell \,N_f\,N\,,
\ee
as the energy levels below $\mu$ get filled.

\begin{figure}[h]
\begin{center}
\epsfig{file=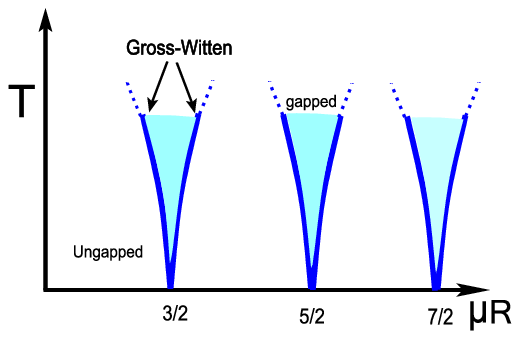, width =2.7in}\hspace{0.5in}
\epsfig{file=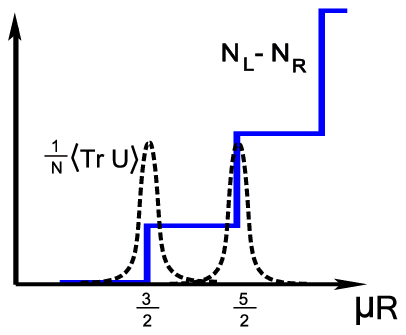, width =2.1in}
\end{center}
\caption{{\underline{Left}}: The low $T$, phase structure at zero coupling and 
large $N$ with  fixed $N_f/N$ on $S^3$. {\underline{Right}}:
Behaviour of $N_L-N_R$ and the Polyakov loop 
with increasing $\mu$ at low $T$.
}
\label{number}
\end{figure}
\subsection{Weak (zero) coupling with $N_f/N$ fixed}
The effect is particularly intricate in the Veneziano limit (assuming
for the present purposes that the axial $U(1)$ 
symmetry is anomaly free) on the three-sphere. At non-zero
temperatures, the Euclidean theory on $S^3 \times S^1$ reduces to a
unitary matrix model for the Polyakov loop matrix 
$U\equiv \exp i\oint_{S^1} A_0$. This effective theory is obtained by
integrating out all massive Kaluza-Klein modes on $S^3$ 
(e.g \cite{Aharony:2003sx}). As is well known \cite{Aharony:2003sx}
from this analysis, the free theory without fundamental matter
experiences a Hagedorn/deconfinement transition at large-$N$. In
particular, at low temperatures the theory is in a confined phase with
$\frac{1}{N}\langle {\rm Tr} U\rangle =0$, while beyond the
deconfinement transition this order parameter acquires a non-zero
expectation value.

Incorporating a large number $N_f$  
of fundamental matter fields with $N_f/N$ fixed, 
changes the low temperature picture, 
especially when a chemical potential for the fermions is
introduced. The situations with baryon number chemical potential 
\cite{Hands:2010zp} and $U(1)_R$ chemical potential (see Appendix \ref{appa})
for the fermions are quite similar, although the respective 
technical details differ.

From the analysis in Appendix \ref{appa}, we see that as the flavour R-charge
chemical potential $\mu$ is increased towards the energy level
$\vep_\ell$ of a fermion mode, the theory experiences a third-order
Gross-Witten transition \cite{Gross:1980he} at 
\be
\mu_-^* \,\simeq\, \vep_\ell - T\, \ln\left(\tfrac{N_f}{N}\,d_\ell\right)\,.  
\ee
Here, the low temperature approximation implies that $\vep_\ell
\gg  T$. At this transition, the large-$N$ distribution of the  
eigenvalues of the unitary matrix $U$ 
changes from an ungapped to a
gapped phase. The distribution, however, quickly reverts to its
ungapped phase as $\mu$ is increased slightly past $\vep_\ell$, at 
\be
\mu_+^* \,\simeq\, \vep_\ell + T\, \ln\left(\tfrac{N_f}{N}\,d_\ell\right)\,
\ee
which represents yet another Gross-Witten transition. This
pair of transitions is accompanied by 
two qualitatively interesting effects. The first is a change in the 
Polyakov loop
order parameter $\frac{1}{N}\langle {\rm Tr}U\rangle$ which is small
in the ungapped phases, but becomes large in the gapped phase. The
second physical effect is the chiral fermion number which jumps across
the two transitions so that $\Delta(N_L-N_R)=d_\ell N N_f$. These
phenomena are depicted in Figure \eqref{number}. The work of
\cite{Hands:2010zp} has shown that these qualitative 
effects (with a baryon number
chemical potential) persist in QCD ($N=3$) in a small volume.

\subsection{Green's function for $\tilde\psi_q \psi_q$ at weak coupling}
The study of the large-$N$ theory at strong coupling and 
with $N_f/N$ fixed is beyond the
scope of this paper. The main point of the weak coupling picture
described above is to indicate the various effects associated to
threshold values of the chemical potential. As we will see below, the
elementary fermionic states are not visible at strong
coupling; instead, the fluctuations of probe D-branes  at strong
coupling comprise of composite gauge invariant
operators made up of fundamental fermions and their superpartners. The
simplest amongst these is the fermion bilinear
\be
{\cal O}\,\equiv \sum_{i=1}^{N_f}\tilde \psi_{q\,i}\psi^i_q\,.
\ee
This is a dimension three operator at weak coupling. In the context of
the D3-D7 system its conformal dimension is protected by its R-charge
\cite{karchkatz}.  However, the normalization of its correlator can and
does receive quantum corrections as a function of
$\lambda=g^2_{YM}N$. At strong coupling, in the presence of an
R-charge chemical potential, we will be interested in computing the
two-point function of this operator on $S^3$. It will be extremely
interesting to contrast this with perturbative results.
 \begin{figure}[h]
\begin{center}
\epsfig{file=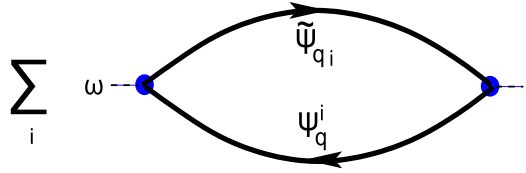, width =2.7in}
\end{center}
\caption{ Two-point correlator of $\tilde\psi_q\psi_q$ in the free
  theory, in frequency space.}
\label{twopt}
\end{figure}
The frequency space retarded 
correlator of ${\cal O}$, at zero external spatial
momentum is, 
\be
\tilde{\cal G}_R(\omega) \,=\,i\int_{0}^\infty dt
\, e^{i\omega t}\int 
\frac{d^3x}{\rm Vol}\, 
\langle \left[{\cal O}(\vec x, t)\,,{\cal O}(0)\right]\rangle\,.
\ee
Using the finite density propagators \eqref{finitemu}
for the internal fermions in Fig.\eqref{twopt}, we obtain a general
form for this correlator at zero temperature and non-zero chemical
potential $\mu>0$,
\be
\tilde{\cal G}_R(\omega)\,=\,\frac{N_f N}{2\pi^2 R^3} \sum_{\ell=1}^\infty d_\ell\,\left(\frac{1}{\omega+2\vep_\ell}-\frac{1}{\omega-2\vep_\ell}\right)
\left[1-2\Theta(\mu-\vep_\ell)\right]\,.
\ee
For the retarded correlator we need to choose an ``$i\epsilon$''
prescription where the poles in the correlator are taken to be
slightly below the real axis.
The simple poles in $\omega$ represent on-shell intermediate modes
with vanishing external spatial momentum. The step function
$\Theta(\mu-\vep_\ell)$ is the Fermi-Dirac distribution evaluated 
at zero temperature, and exhibits the ``Pauli-blocking'' of
occupied levels below the energy $\mu$. Using the energy levels
$\vep_\ell$ and degeneracies for $S^3$, the Green's function can
be written in terms of digamma functions (up to constants and
``contact'' terms proportional to $\omega^2$)  
\bea
\tilde{\cal G}_R(\omega)\,=\, \frac{N_f N}{16\pi^2 R^2}\,(\omega^2R^2-1)&&
\,\left[\psi\left(\tfrac{1}{2}-\tfrac{1}{2}\omega R\right)+
\psi\left(\tfrac{1}{2}+\tfrac{1}{2}\omega R\right)\right.
\label{weakgf}\\\nonumber\\\nonumber
&&\left.-2\psi\left(\tfrac{3}{2}+n +\tfrac{1}{2}\omega R\right)
-2\psi\left(\tfrac{3}{2}+n -\tfrac{1}{2}\omega R\right)\right]
\\\nonumber
 n\,\equiv\,[\mu\,R-\tfrac{1}{2}]\,.
\eea 
Here $n$ is the largest integer less than or equal to 
$\mu R-\tfrac{1}{2}$. In the
perturbative limit, the chemical potential does not introduce new
poles in the correlator $\tilde{\cal G}(\omega)$. In particular, as expected
there are no instabilities of any kind. In the limit of large $R$,
we should recover the correlator in flat space, after incorporating
the correct $i\epsilon$ prescription,
\bea
\tilde{\cal G}_R(\omega)\,\to -\frac{N_f N}{16\pi^2}\,\omega^2\,
\left[\ln(\omega^2 R^2) +
  \ln\left(\frac{4\mu^2- \omega^2}{\omega^2}\right)^2
+2i\pi\,{\rm sgn}(\omega)\,\Theta(\omega^2-4\mu^2)
\right]
\label{flatweak}
\eea
using $\psi(x)\to \ln x$ for large $x$.
The correlator has 
branch points at $\omega=\pm 2\mu$ and an imaginary part for $\omega^2>4\mu^2$.
This is because states with energy below $|\mu|$ are occupied 
and therefore Pauli-blocked. Only when the external energy $|\omega|$
is larger than $2|\mu|$, can the intermediate states be on-shell and
yield an imaginary part for the Green's function.
When $\mu=0$, the correlator has the expected behaviour $\sim
\omega^2\ln\omega$ for a dimension three operator. In particular, at
$\mu=0$ we expect the two-point function (including when $R$ is finite)
to have the same form for all
values of the 't Hooft coupling , although the normalization can
and does receive quantum corrections. 
We now turn to the strong coupling description of the system in the probe
limit $N_f\ll N$.

\section{Probe D7-branes in Global AdS}
In the large-$N$ limit and at strong 't Hooft coupling
$\lambda=g^2_{YM}N\gg 1$, the flavour modes
correspond to low-energy excitations of 
probe D7-branes in the $AdS_5\times S^5$ 
geometry \cite{karchkatz}. 
The flavour D7-branes wrap an $S^3\subset S^5$ while filling the $AdS_5$
directions.  
Such embeddings of the D7-brane are characterized by the
``slipping angle'' $\theta$, which determines the size of the three-sphere
wrapped by the D7-brane and is defined by 
the following paramerization of the $S^5$ metric
\begin{equation}
d\Omega_5^2=
d\theta^2+\sin^2\theta\,d\phi^2+\cos^2\theta\,d\Omega_3^2\,,\qquad 
0\leq\theta\leq\frac{\pi}{2}\,.
\label{s5}
\end{equation}
A static probe D7-brane wrapping this $S^3$ breaks the $SO(6)$
R-symmetry of ${\cal N}=4$ theory to an $SU(2)_R\times U(1)_R$ subgroup. As
discussed above, the $U(1)_R$ is non-anomalous in the theory with
$\frac{N_f}{N} \to 0$ and is the isometry generated by
shifts of the angle $\phi$.
We will consider probe branes spinning in   
the $U(1)_R$ isometry direction parametrized by the angle $\phi$. The
associated angular velocity is conjugate to the $U(1)_R$ charge
carried by such states. 

When embedded in the global AdS spacetime, the D7-brane world-volume is
topologically $S^3\times S^3 \times {\mathbb R}\times {\mathbb
  R}_t$. The first $S^3$ sits inside the global $AdS_5$ geometry whose
conformal boundary is an $S^3$. The D7-brane wraps a second
$S^3\subset S^5$, specified by the polar angle $\theta$ in
Eq. \eqref{s5}. 
We write the global $AdS_5$ metric in Fefferman-Graham
coordinates \cite{FG}
\be
ds^2=- \frac{(1+ \tfrac{1}{4}\,z^2)^2}{z^2}\, d t^2
+\frac{(1-\tfrac{1}{4}\,z^2)^2}{z^2}\, d\Omega_3^2 +\frac{dz^2}{z^2}\,,
\ee
with $0 < z \leq 2$. The conformal boundary is approached as $z\to
0$. We have set the AdS radius to unity. In these units and the 
coordinate system employed above, the radius of the boundary
three-sphere is $R= 1$.

\subsection{Embedding Ansatz} 
We can take the D7-brane to spin with constant angular velocity in the
$\phi$ direction, while at the
same time allowing for a radial dependence of its angular coordinate
on the $S^5$ \cite{O'Bannon:2008bz, evans, Karch:2007pd}, 
\be
\phi=\,\tilde\mu \,t + \,g(z).
\ee
It turns out (Appendix C) that for massless flavours the
radial dependence must be taken to vanish.
So we  will only consider 
the configurations with $g(z)=0$, as we are mainly interested in the $m=0$
theory when the $U(1)_R$ can be viewed as a symmetry (at least in the
large-$N$ limit).
The angular velocity
$\tilde\mu$ is related  to the R-charge chemical potential in field
theory, as 
\be
\tilde\mu = 2\mu\,.
\ee
The factor of two appears because the two fields excited on the probe
brane,  namely $\theta$ and $\phi$ are related to the magnitude and
phase, respectively, of the fermion bilinear made from flavour
fields. This object has R-charge +2. An R-charge chemical potential
for a fundamental fermion flavour can also be viewed as a time
dependent phase $\exp(i\mu t)$ for the fermion fields. Correspondingly, the
bilinears have a time dependent phase $\exp(2i\mu t)$.

Setting $g(z)=0$ for massless flavours, 
the induced metric on the D7-brane is 
\bea
&&ds^2\Big|_{\rm D7}=\\\nonumber
&& -\left(\frac{(1+\tfrac{1}{4}\,z^2)^2}{z^2}-4\mu^2\sin^2\theta\right)dt^2
+\left(z^2 \theta'(z)^2+ 1\right)\frac{dz^2}{z^2}+ 
\frac{(1-\tfrac{1}{4}\,z^2)^2}{ z^2}d\Omega_3^2+\cos^2\theta d\tilde \Omega_3^2.
\eea
The D7-brane probe dynamics is then controlled by the DBI action
\be
S_{\rm D7} = \,2\pi^2{\cal N}_{\rm D7}\,
\int dt\,dz \, \cos^3\theta(z)\,\frac{(1-\tfrac{1}{4}\,z^2)^3}{z^5}
\sqrt{(1+z^2\theta'(z)^2)((1+\tfrac{1}{4}\,z^2)^2 - 4\mu^2 \,z^2
  \sin^2\theta)},
\label{dbi1}
\ee
where ${\cal N}_{\rm D7}$ depends on the D-brane tension $T_{\rm D7}$ and the
number of flavours $N_f$, as
\be
{\cal N}_{D7}\, \equiv  2\pi^2\,N_f \, T_{\rm D7} = \frac{\lambda}{2(2\pi)^4} N_f N\,. 
\ee
The additional normalization factor of $2\pi^2$ in
Eq. \eqref{dbi1}, can be traced to  
the volume of the boundary three-sphere, whose radius is
$1$ in our units.

Classical solutions to the equations of motion resulting 
from varying Eq. \eqref{dbi1}
have the following general behaviour near the AdS boundary,
\be
\theta(z)\,|_{z\to 0} = \theta_{(0)}\, z + \theta_{(2)}\, z^3 +
\tfrac{1}{2}\,\theta_{(0)}(1-4\mu^2)\,z^3\ln z + \ldots.
\label{asymp}
\ee
From the standard AdS/CFT dictionary \cite{magoo}
this  is the behaviour expected for a
supergravity mode dual to a dimension 3 operator in the field
theory. It is interesting to note that in global AdS, there is an
extra logarithmic term in the asymptotic expansion which explicitly
depends on the chemical potential. This will give rise to a chemical
potential dependent counterterm for the probe brane action below. 
The operator must be made from flavour fields and must
transform under the global $U(1)_R$ symmetry. 
It is natural then, to associate the slipping angle $\theta$ with the 
quark bilinear\footnote{ The precise mapping relates $\theta$ to
  the operator $\tilde \psi_{q\,i}\psi_q^i +
  q_i^\dagger\Phi_3\,q^i+\tilde q_i\Phi_3\,\tilde q^{\dagger \,i}$.}
\begin{equation}
{\cal O} = \sum_{i=1}^{N_f}\,\tilde{\psi}_{q\,i}\, \psi_q^i.  
\end{equation}
The leading coefficient $\theta_{(0)}$ , when non-zero, results in a bare quark
(hypermultiplet) mass which
explicitly breaks the $U(1)_R$ symmetry. For vanishing hypermultiplet mass, the
coefficient of the $z^3$ term directly yields the quark condensate
$\langle {\cal O}\rangle$. A non-zero quark condensate breaks the
$U(1)_R$ symmetry spontaneously (assuming that the quark mass is zero). The
precise dictionary between the gauge theory parameters and
supergravity modes is quoted below.

\subsection{Divergences and renormalization}
The D7-brane probe action is formally divergent and needs to be
regulated and renormalized. The regularization of
these divergences is well understood within the general framework of
the holographic-RG approach applied to probe D-branes 
\cite{karchskenderis,andythesis,Karch:2009ph}. It is instructive to
first look at all the divergent counterterms  implied by the 
near-boundary asymptotics \eqref{asymp} in the D7-brane action, and
then compare with the results of the holographic RG method. By cutting
off the $z$ integration at $z=\epsilon \ll 1$, we find the following 
divergent terms 
\be
S_{\rm D7}^{\,\epsilon} \simeq 2\pi^2\,{\cal N}_{\rm D7}\,\int dt\,
\left(- \frac{1}{4\epsilon^4} +
  \frac{1}{4\epsilon^2}+\frac{\theta_{(0)}^2}{2\epsilon^2}
+  \theta_{(0)}^2\,(1-4\mu^2)\,\ln\epsilon+\ldots\right)\,,
\ee 
Note that the chemical potential $\mu$ contributes to the
logarithmically divergent counterterm. 
However, when $\theta_{(0)}=0$ and the hypermultiplets are
massless, the divergences are $\mu$-independent. This is consistent
with the general expectation that much like a finite temperature, a
chemical potential should not introduce new divergences in
the theory. For non-zero hypermultiplet masses, $U(1)_R$ is not a good
symmetry and the angular velocity $\mu$ leads to a time-dependent
mass and also makes an appearance in the UV divergent terms. 

The divergences are removed by the addition of local counterterms
at $z=\epsilon$ to yield the regularized action
\be
S^{\rm reg}= S_{\rm D7}^{\,\epsilon} + {\cal N}_{\rm D7}\,
\int d\Omega_3\,dt\,\sum_{k=1}^4 L_k
\label{regaction}
\ee
Denoting the boundary metric of the cut off AdS space as $\gamma_{\mu\nu} 
dx^\mu dx^\nu\equiv 
\tfrac{1}{\epsilon^2}(- dt^2+ d\Omega_3^2)$, 
we can write these counterterms in terms of the Ricci scalar
${\cal R}$ of the boundary at $z=\epsilon$ 
as in \cite{karchskenderis,andythesis, Karch:2009ph}:
\bea
&&L_1= - \frac{1}{4}\sqrt{-\gamma}\,,\qquad L_2 = 
\frac{1}{24}\sqrt{-\gamma}\,{\cal R}\,,\qquad L_3= 0\,,\qquad L_4=
\frac{1}{2}\sqrt{-\gamma}\,\theta(\epsilon)^2\qquad\\\nonumber  
&&L_5=\sqrt{-\gamma}\,(\,\tfrac{1}{6}{\cal
  R}+4\mu^2\,\gamma^{tt})\,\theta(\epsilon)^2\ln \theta(\epsilon)\,,\qquad
L_6= - \frac{5}{12}\sqrt{-\gamma}\,\theta(\epsilon)^4\,. 
\eea
We have used the fact that $\theta(\epsilon)$ is given by
Eq.\eqref{asymp} as 
$\epsilon \to 0$ and that  ${\cal R}= 6 \epsilon^2$. Notice that the
counterterm $L_5$ depends on the chemical potential $\mu$. The form of
this counterterm can be deduced by first noting that that $\theta$ and
$\phi$ can be packaged into a complex scalar $\Phi$ and the
counterterms for this scalar can be expressed in the form of Eq.(4.2)
in \cite{karchskenderis}
\footnote{I would like to thank Andy O' Bannon for pointing this out to
  me.}.

Having defined the regulated action, the 
dictionary between the gauge theory parameters, namely the quark mass
and condensate, and corresponding supergravity modes 
can be worked out easily (having set the AdS radius to
unity). Specifically, for massless hypermultiplets we obtain
\be
m= \frac{\theta_{(0)}}{2\pi\alpha'}\,=0\,,\qquad\quad
\langle{\tilde{\psi}_{q}\psi_q}\rangle = {\cal N}_{\rm D7}\,\theta_{(2)}
\label{asymp}
\ee

\subsection{Solutions}
The DBI equations of motion have a large family of spinning
solutions. The solutions fall into two topologically distinct
varieties. In one class of solutions, the $S^3\subset S^5$ which is 
wrapped by the D7-brane probe, shrinks in the interior, before the
brane reaches the centre of the global AdS space. This type of
solution always yields a non-zero  hypermultiplet mass  which can be
read off from the asymptotic behaviour of $\theta(z)$. The second
category of solutions are those where the second $S^3$ embedded in $AdS_5$
shrinks, and the probe brane reaches the origin of AdS. A very
  interesting topology-changing, but continuous phase transition
  between these two classes of solutions was discovered in 
\cite{Karch:2009ph,karch1}. This phase transition was originally found
and explained at zero
temperature, at a critical value of the hypermultiplet mass. It can be
thought of as a finite volume, meson binding/unbinding transition at
strong coupling and zero temperature.   
The second category of solutions - the AdS-filling embeddings, 
include configurations corresponding to massless hypermultiplets. In
this phase, a maximally separated static quark-antiquark pair is
stable against light quark pair production.
We will be interested primarily in the AdS-filling massless embeddings
which do not explicitly break the $U(1)_R$ symmetry. 

\subsubsection{Constant Solution} 
The simplest solution to the DBI
equations of motion following from Eq. \eqref{dbi1} is
$\theta(z)=0$. This has $m=\langle{\cal O}\rangle =0$. The free energy
or the action per unit time for
the constant solution is 
\be
F = 2\pi^2\,{\cal N}_{\rm D7}\, \left[\int_\epsilon^2
dz \,\frac{1}{z^5}(1-\tfrac{1}{4}\,z^2)^3(1+\tfrac{1}{4}\,z^2)
-\frac{1}{4\epsilon^4}+\frac{1}{4\epsilon^2}
\right] \,=\,\frac{3\pi^2}{16}\,{\cal N}_{\rm D7}\,.
\ee
The induced metric on the D7-brane is simply (global) $AdS_5\times
S^3$. The charge density conjugate to $\mu$
\be
d \equiv {\cal N}_{\rm D7}\int_0^2 dz \,\frac{\partial \cal L}{\partial \mu}
\ee
evaluates to zero for the $\theta=0$ solution.

The trivial solution with $m=0$ and zero R-charge density ($\theta(z)=0$) 
exists for all values of $\mu$, but as we will
see below, it is not always thermodynamically stable.

\subsubsection{Non-constant solutions} 
We will now examine in detail the possibility that 
there may be additional, non-constant solutions to the DBI
equations of motion with $m=\theta_{(0)}=0$. If such embeddings were
to exist, they would correspond to configurations with a finite
R-charge density and which also spontaneously break the $U(1)$
R-symmetry by the formation of a fermion bilinear condensate
$\theta_{(2)}\neq 0$.

To find out
whether such configurations exist, given the non-linearity of the DBI
action and equations of motion, we first employ a brute force
numerical approach. For non-constant solutions which fill AdS space 
\footnote{Solutions which cap off before getting to the centre of AdS
  spacetime, always correspond to massive flavours, and fall in the
  category of so-called ``Minkowski embeddings''.}, 
smoothness at the origin of $AdS_5$ requires
\begin{equation}
\theta(z)\big|_{z=2} = \Upsilon = {\rm constant}\,;\qquad 
\theta'(z)\big|_{z=2}=0\,.
\end{equation}
Furthermore, reality of the DBI action \eqref{dbi1} imposes the restriction 
\be
\sin\theta(z)\big|_{z=2}\leq \frac{1}{2\mu} \label{restrict}
\ee
so that the action is real. At this point we stress that we only focus
attention on the solutions having $\phi=2\mu\,t+g(z)$ with
$g(z)=0$. This is because we want to isolate the $m=0$ embeddings and
for these $g(z)$ must necessarily vanish, as argued in Appendix C. 
For large enough $\mu$, we
may expect $\theta(z)$ to be small everywhere, and apply a linearized
approximation to the equation of motion. We will return to this
shortly. 

The full space of non-trivial spinning D7-brane embeddings for general
$\mu$, includes both the AdS-filling solutions and the solutions which
cap off before getting to the orgin of $AdS_5$. The latter class of
embeddings are characterized by the fact that the $S^3 \subset S^5$
wrapped by them shrinks smoothly before the brane gets
to the centre of $AdS_5$:
\be
\theta(z_0) =\frac{\pi}{2}\,,\qquad\theta'(z_0)\to \infty\,,\qquad
{\rm for \,\,\, some}\,\,z_0 < 2.
\ee

\begin{figure}[h]
\begin{center}
\epsfig{file=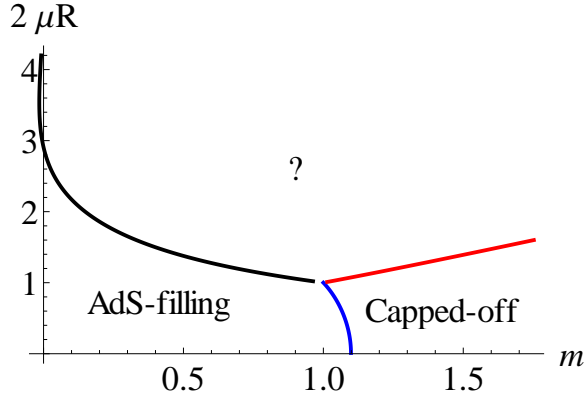, width =3.0in}
\end{center}
\caption{ The space of known spinning D7-brane embeddings in global
  $AdS_5\times S^5$. The known solutions all lie to one side of the
  curves shown. The figure also shows that near $\mu R \approx
  \frac{3}{2}$,
  there is a solution with $m=0$. (Note that we have reinstated the
  radius $R$ of the boundary sphere, which was set to $1$ by our
  metric conventions).
}
\label{d7phase}
\end{figure}

By numerically solving the DBI equations of motion for different
values of $\mu$, we extract the mass parameter $m$ and condensate 
$\langle\tilde\psi_q\psi_q\rangle$ from the asymptotics \eqref{asymp}.
In this way we can obtain all possible types of D7-brane embeddings
which have the slipping angle $\theta(z)$ as the only non-trivial
mode. The possible space of solutions is summarized in Fig.\eqref{d7phase}

\begin{figure}[h]
\begin{center}
\epsfig{file=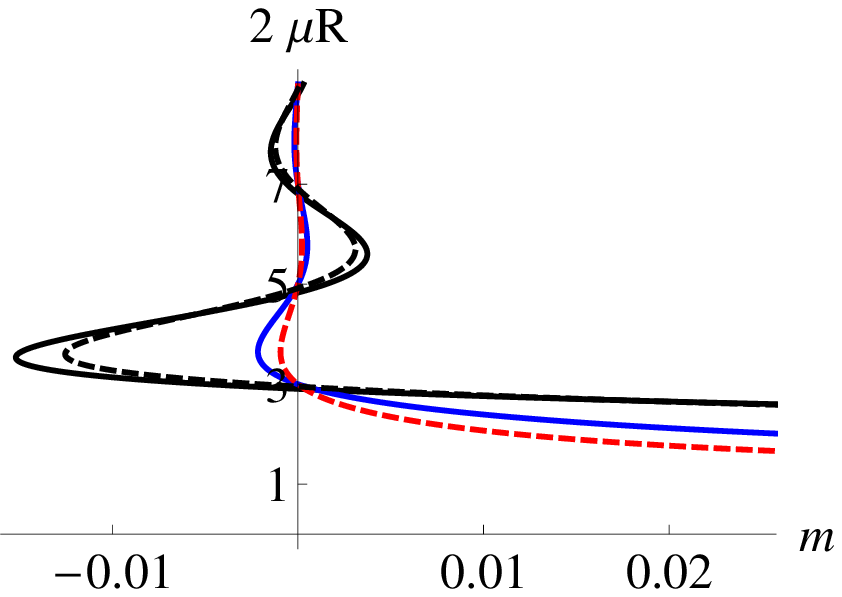, width =2.5in}
\hspace{0.5in}
\epsfig{file=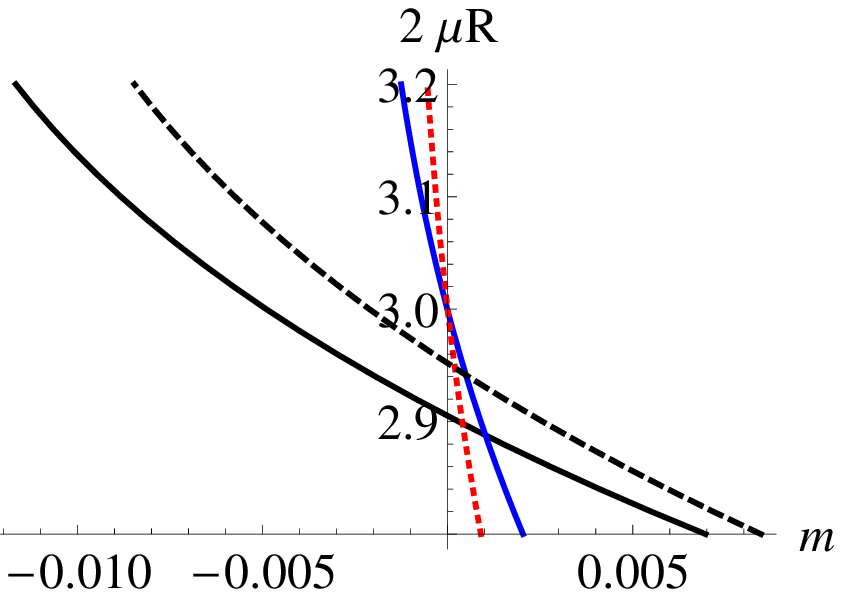, width =2.5in}
\end{center}
\caption{ 
{\underline{Left}}: 
A blown-up plot of 
the masses and chemical potentials for three families of AdS-filling solutions. 
The black line is the boundary line depicted in
Fig. 4. New massless embeddings appear in the vicinity of
$\mu R =\frac{3}{2},\frac{5}{2},\frac{7}{2}\ldots$. {\underline{Right}}: Closer examination reveals
massless $(m=0)$ 
finite density solutions for a finite range of $2\mu R$, close to
the odd integers.
}
\label{blowup}
\end{figure}

The figure exhibits three ``phase boundaries'':
\begin{itemize}
\item{ The line in blue separates
``AdS-filling''
solutions from the ``capped-off'' embeddings. This boundary line is
obtained by looking at AdS-filling embeddings with
$\theta(z)\big|_{z=2}\to\pi/2$ and $\theta'(z)\big|_{z=2}=0$. 
These are on the verge of
getting capped off and form a phase boundary for a continuous
topology-changing transition of the type 
found originally in \cite{karch1,Karch:2009ph}.}

\item{The red line is the maximum allowed value of the chemical
    potential $\mu$, below which the DBI action still remains
    real, for the massive capped-off solutions or so-called 
``Minkowski embeddings''. This is obtained by
    noting that a D7-brane ending at $z=z_0 < 2$ with
    $\theta(z_0)=\pi/2$ must have $(1+\tfrac{1}{4}z_0^2) \geq 2 \mu  z_0$ for
    every allowed value of mass (from \eqref{dbi1}).}

\item{The third boundary line in black is of interest to us and it
    represents solutions that fill $AdS_5$, with the maximum allowed
    value of $\mu$ dictated by reality of the DBI action \eqref{dbi1}.
    These saturate the inequality \eqref{restrict}, so that
    $2\mu=1/\sin\theta(z)|_{z=2}$. A key feature of this family of 
    solutions is that it
    intersects the $m=0$ axis at several points close to (but not
    precisely at) odd integer values of $\mu R$, (Fig. \eqref{blowup}). } 
\end{itemize}

Blowing up the region for small $m$, we see clearly in
Fig. \eqref{blowup}, that there exist non-trivial solutions with
$m=0$. By plotting the masses for four families of solutions, we see
that the non-constant, vanishing mass embeddings occur near 
$2 \mu R = 3, 5,7, \ldots$.  It is also clear from the
plots that there is a finite range of values of $2 \mu R$ in the vicinity
of the odd-integer points, for which non-trivial massless solutions
arise. For example, it seems that such finite density massless 
solutions to the DBI equations of motion appear for $2.9 \lesssim 2 \mu R
\lesssim 3$. 

The four families of solutions, whose masses are displayed in
Fig.\eqref{blowup}, were obtained by integrating the equations of
motion outwards from the origin ($z=2$) with boundary
conditions, assuming that $2\mu >1 $:
\bea
&&\theta'(z)\big|_{z=2}=0\,,\qquad \theta(z=2) =
\sin^{-1}\left(\frac{1}{n(2\mu-1)+1}\right)\qquad n =1\,,
2\,,15\,,35\,.
\nonumber
\label{bc}
\eea
The values of $n$ were chosen to illustrate the full range of $\mu$ in
Fig.\eqref{blowup}, in
the vicinity of $2\mu R =3$, for which new $m=0$ embeddings exist.
The limiting case for which the DBI action still remains real,
corresponds to $n=1$ (solid black curve). The families labelled by
$n=2, 15$ and $35$, all satisfy Eq.\eqref{restrict} and are
represented by the dotted-black, solid-blue and dotted-red curves,
respectively in Fig.\eqref{blowup}.

\begin{figure}[h]
\begin{center}
\epsfig{file=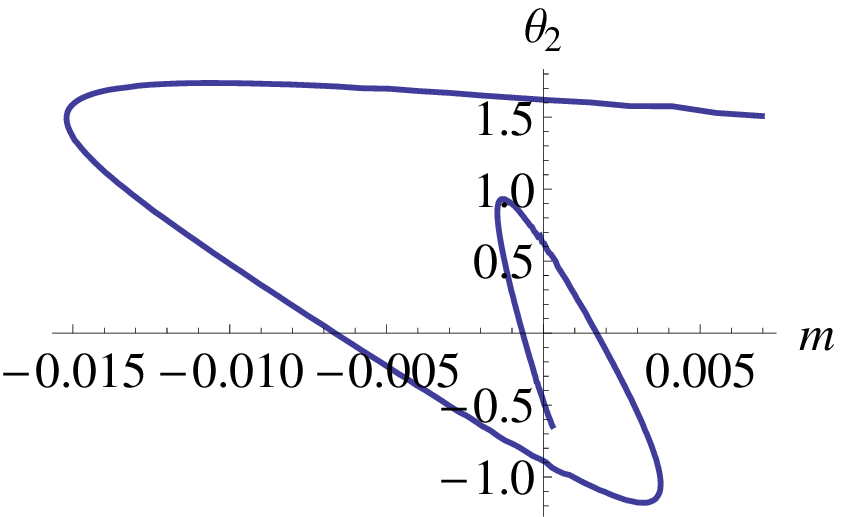, width =2.5in}
\hspace{0.5in}
\epsfig{file=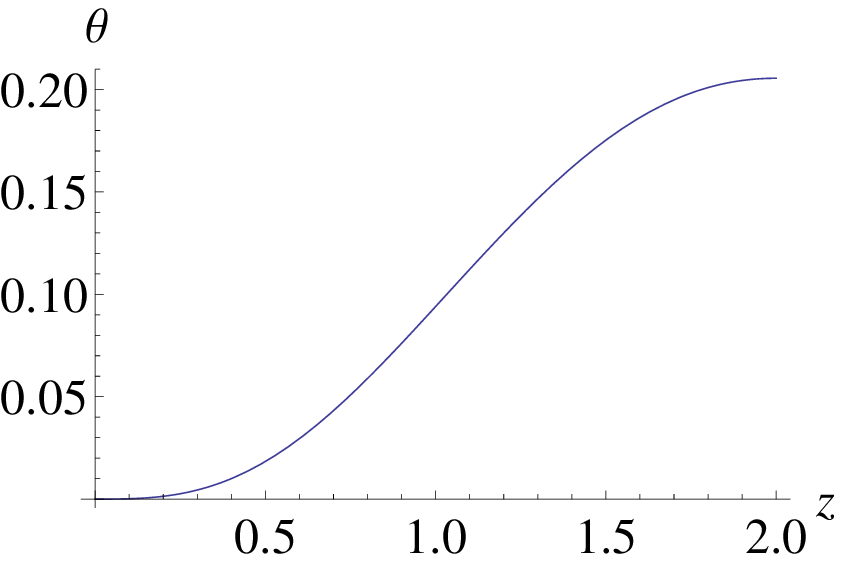, width =2.5in}
\end{center}
\caption{ {\underline {Left}}:
The normalizable mode $\theta_{(2)}$ 
which is proportional to the chiral condensate 
$\langle \tilde\psi_q\psi_q\rangle$ when $m=0$. These are the values
extracted from the $n=1$ family of solutions. 
{\underline{Right}}: 
The non-constant massless solution for $2\mu R \approx 2.95$.
}
\label{sample}
\end{figure}

All the configurations above (including those with $m=0$) will
necessarily have a non-vanishing chiral condensate $\langle \tilde \psi_q
\psi_q\rangle$, indicating that the solutions with $m=0$, dynamically
break the chiral $U(1)_R$ symmetry (see Fig.\eqref{sample}). 
What remains to be established is
whether these configurations are stable and whether they are 
thermodynamically favoured over the constant $\theta =0$ solution.

\subsubsection{Free energy and saddle points}

We now turn to the key issue, namely, how the free energies of the 
non-trivial D7-brane embeddings discussed above compare with that of the
constant $\theta=0$ embedding. We first note that, as $\mu R$ is
dialed from zero, $\theta=0$ remains the only solution of the DBI
equations of motion. A non-constant embedding first appears  at $2\mu R
\approx 2.9$ (from the $n=1$ family above).

Using the regulated form of the action \eqref{regaction}, 
we numerically evaluate on
each of the $m=0$ solutions, the difference of the regulated actions, 
\be
\Delta S \,\equiv \,
(2\pi^2 {{\cal N}_{\rm D7}})^{-1}\,\left[\,S^{\rm reg}(m=0,\mu R) - S^{\rm
    reg}\big|_{\theta=0}\right]\,. 
\ee
In all cases we find that the non-constant embeddings in the window 
$2.9\lesssim 2\mu R\lesssim 3$ have a higher action than the
$\theta=0$ solution. Thus these must either be unstable or metastable.
In addition, in the limit that $2\mu R \to 3$, the non-constant
embeddings approach $\theta=0$. 

\begin{figure}[h]
\begin{center}
\epsfig{file=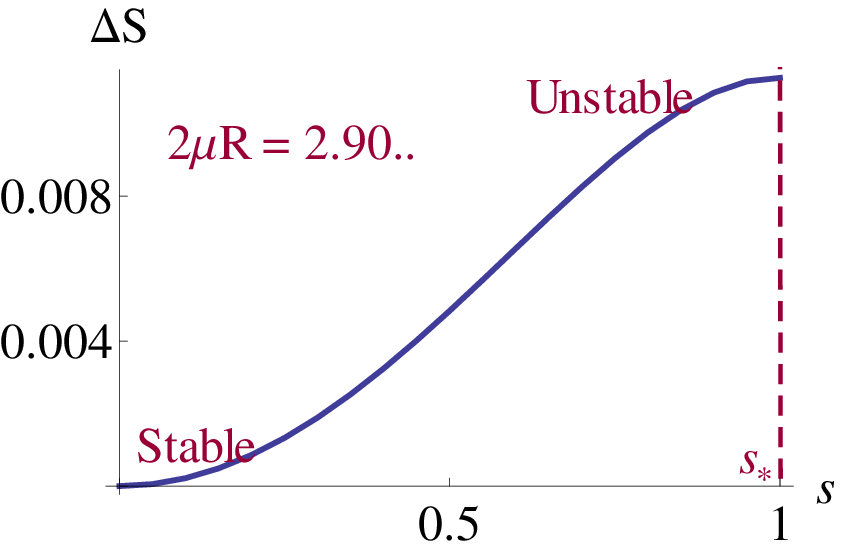, width =2.5in}
\hspace{0.5in}
\epsfig{file=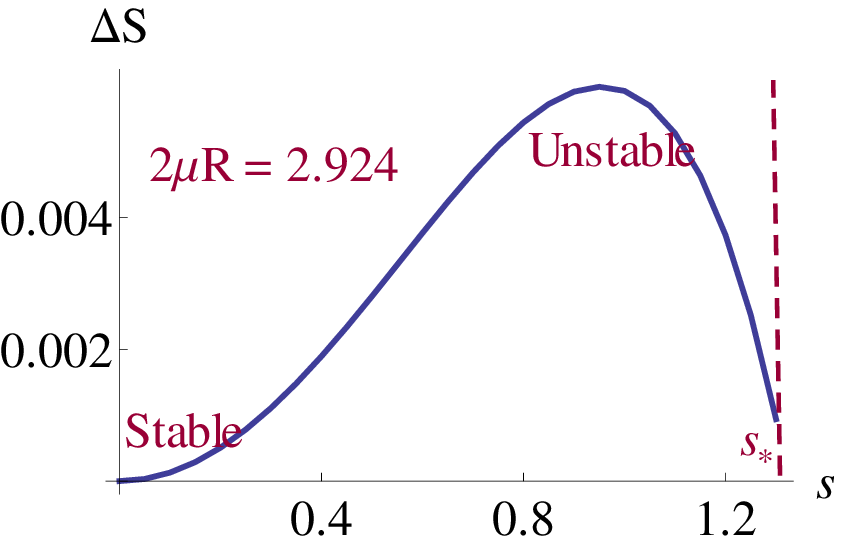, width =2.5in}
\end{center}
\hspace{0.5in}
\begin{center}
\epsfig{file=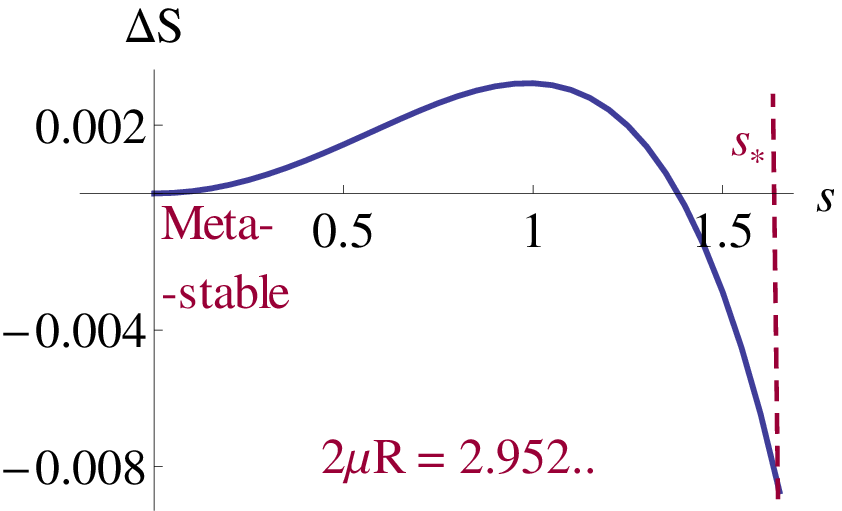, width =2.5in}
\hspace{0.5in}
\epsfig{file=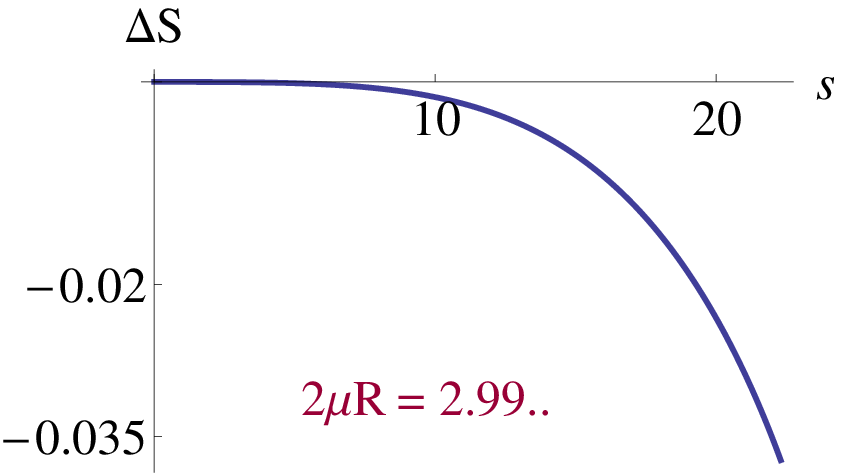, width =2.5in}
\end{center}
\caption{ \small{$\Delta F$, the difference in free energy (action per
    unit time) between non-constant 
massless configurations and the zero density state, evaluated along
a one-parameter family of  
configurations labelled by $s$. 
The point $s=0$ represents $\theta(z)=0$, while $s=1$  corresponds to
the new non-constant solution which first appears when $2\mu R\gtrsim 2.9$. 
The four plots illustrate what happens when $\mu
  R$ is increased. At $2\mu R=3$, the critical point, the two saddles merge and 
the system has a runaway instability.}}
\label{saddle}
\end{figure}

The complete situation is best described by the following procedure leading to 
Fig.\eqref{saddle}. For every non-trivial solution $\theta(z)$, 
obtained with boundary conditions of the form \eqref{bc}, we define a
one-parameter family of configurations which {\em do not} necessarily 
solve the DBI
equations of motion,
\bea
&&\theta (z ; s )\,\equiv \,
s\,\theta(z)\big|_{m=0,\mu\neq 0}\, \qquad 0\leq s \leq s_* \,,
\\\nonumber\\\nonumber
&&s_* \,\theta(z)\big|_{z=2}\equiv \,\sin^{-1}\left(\frac{1}{2\mu}\right)\,.
\eea
This is a simple rescaling of a given non-trivial solution, where the
rescaling parameter is bounded by the requirement that the DBI action
evaluated on the configuration be real.
It
yields a one dimensional slice of the
configuration space of $\theta(z)$ (with the condition that the 
hypermultiplet mass evaluates to zero at
the boundary), along which, only the two points $s=0$ and $s=1$
represent  saddle point
configurations or solutions of the DBI equations of motion 
for a fixed $\mu$. It is important that $s$ is bounded by some value
$s_*$ beyond which the DBI action evaluated on these rescaled
configurations ceases to be real. Hence the configuration space of
$\theta(z)$ has a boundary for some fixed $\mu$. 

In Fig. \eqref{saddle} we plot the
action evaluated on these configurations and find that the
non-constant solutions are always unstable. In contrast, the
zero density, $\theta=0$ embedding is stable for low $\mu$, but becomes
{\em metastable} as $\mu$ is increased, eventually becoming {\em unstable} at
$2\mu R\approx 3$. The end-point of the metastability/instability
appears to lie outside the region of validity of 
the ansatz for our probe D-brane setup.
The main result we take away from this analysis is that the $\theta=0$
solution becomes metastable and eventually encounters a runaway
instability at $2\mu R= 3$. 

The entire picture described in this section is repeated 
for every odd-integer value of $2 \mu R$. 
The unstable direction at each of these critical points is towards 
a non-constant
massless embedding $\theta(z)$. The critical values $2\mu R = 2\ell
+1$, indicate that the modes condensing are fermion {\em bilinears}
which have R-charge 2 and are made from 
elementary fermion harmonics carrying energy
$(\ell+\tfrac{1}{2})R^{-1}$. The mass to charge ratio of the bilinears
is $(\ell+\tfrac{1}{2})R^{-1}$ and governs the critical value of $\mu
R$ for the onset of a given instability.
Schematically, the operator
$\tilde\psi_{q\,i}\psi_q^i$ can be expressed as a sum over spherical
harmonics, whose homogeneous mode on $S^3$ has the form,
\be
\frac{1}{2\pi^2}
\int d^3\Omega \,\,\tilde\psi_{q\,i}\psi_q^i \,=\, \sum_{\ell,\,n} 
\tilde\psi_{\ell\,,n}^i\,\psi_{\ell\,,-n }^i,
\ee
with $\psi_{\ell\,,n}$ an eigenmode of the Dirac operator on $S^3$ 
with mass $(\ell+\tfrac{1}{2})R^{-1}$, $\ell=1,2,\ldots$ and $n$
denotes the azimuthal quantum numbers associated to a given harmonic.
Therefore, the
instabilities are triggered when the chemical potential times the
R-charge exceeds the
mass of the composite charged scalar 
${\cal O}_\ell\sim \sum_n\tilde\psi_{\ell\,,n}\psi_{\ell\,,-n}$. 

The unstable directions seen above 
 are also qualitatively distinct from the runaway
behaviour encountered in ${\cal N}=4$ SYM for large enough R-charge
\cite{Yamada:2006rx, Yamada:2007gb, Hollowood:2008gp} where the
effect is due to the presence of flat directions for R-charged
scalars. In that case, the scalar ``Coulomb branch'' instability is
 also visible at weak coupling.
In the present setup, the (axial) R-charge 
is carried by the flavour fermions only. In the weakly interacting
theory, when the chemical potential hits half-integer values, the
chiral fermion 
number jumps and there is no instability (see Section\eqref{weak}).
In contrast, the elementary fermions are not visible in the 
strongly interacting limit; instead condensates of fermion bilinears
 are preferred when the chemical potential approaches the same
 half-integer points. At strong coupling these values are naturally
 interpreted as the mass to charge ratios of fermion bilinear modes.
Surprisingly, the DBI
 action, at least within our ansatz, does not appear to include the
 end-point of this instability. Below we speculate on possible resolutions
 of this.

\section{Analytical results: instabilities}
We now turn to an analytical approach to reveal the origin of
the instabilities we have encountered above, and in the process gain
some understanding of the nature of the numerical solutions. 
The main features of the numerical
approach, such as the appearance of new albeit unstable classical solutions,
are a consequence of the non-linearities of the DBI
action. However, we expect some quantitative aspects to be accessible
in a linearized regime, especially for large $\mu$. This can be
understood to be a consequence of Eq.\eqref{restrict}, which requires
that for large enough $\mu R$, the amplitude of $\theta(z)$ is
essentially bounded by $1/(2\mu R)$. The metastabilities and
instabilities we saw above appear at an infinite set of values
$2\mu R=3,5,7,\ldots$, and so we expect 
some of this physics to be accessible in the
linearized regime.

\subsection{Non-constant massless embeddings}
The first issue we would like to understand, is the appearance of a
new saddle point or solution to the  equations of motion with a
finite density. 
When $\mu\gg1$, $\sin\theta(z)< (2\mu)^{-1}\ll 1$, and it is then consistent
to linearize the DBI equations of motion:
\be
\theta''(z) -\frac{48+16 z^2+ 5 z^4}{z(16-z^4)}\,\theta'(z)+
\frac{48+3z^4+ 8 z^2(3+8\mu^2)}{z^2(4+z^2)^2}\, \theta(z)=0.
\label{lineareq}
\ee
As we will see more explicitly in the next section when we compute the
two-point function for fluctuations around the $\theta=0$ solution,
this is the equation of motion for a charged scalar coupled to a
constant background gauge field in global AdS spacetime.
It turns out that the equation can be exactly solved in terms of
hypergeometric functions. This is somewhat easier to see after the
change of variables
\be
z= 2\sqrt y\,,\qquad \varphi(t)\equiv \frac {\theta(\sqrt y)}{f(y)}\,,\qquad
f(y)\equiv (y-1)\sqrt{y- y^{-1}}\,,
\ee
following which we obtain a Schr\"odinger equation for $\varphi(y)$,
\\
\be
\frac{d^2\varphi}{dy^2}+
\left(\frac{(4\mu^2-1)(y^2+1)}{y(y^2-1)^2}-\frac{8\mu^2+1}{(y^2-1)^2}\right)
\varphi =0. 
\ee
\\
The equation is solved by regularized hypergeometric functions
\\
\bea
\theta(z) &= &C_1 \,\sqrt{y} \tfrac{(1+y)^{2\mu}}{(1-y)^{2\mu+1}}
\,\,{}_2\tilde
F_{1}\left(\mu-\tfrac{1}{2},\,\mu+\tfrac{1}{2};2\mu+1 ;\,\left(
\tfrac{1+y}{1-y}\right)^2\right)\\\nonumber  
&+& C_2 \,\sqrt{y} \tfrac{(1-y)^{2\mu-1}}{(1+y)^{2\mu}}
\,\,{}_2\tilde
F_{1}\left(-\mu -\tfrac{1}{2},-\mu+\tfrac{1}{2};\,1-2\mu ;\,\left(
\tfrac{1+y}{1-y}\right)^2
\right)\,,
\eea
\\
where the coefficients are determined by requiring
regularity at the origin of AdS spacetime at $z=2$ or $y=1$. 
Specifically, we are interested in embeddings that fill $AdS_5$
and get to the origin of the space, satisfying
\be
\theta(z)\big|_{z=2}={\Upsilon}\,,\qquad\qquad\qquad\theta'(z)\big|_{z=2}=0.
\ee
Imposing these boundary conditions we find
\be
C_1= i\,\frac{4 \Upsilon e^{i\pi \mu}}{2\mu-1} \,
\frac{\Gamma\left(\tfrac{1}{2}+\mu\right)^2}
{\pi \tan\left(\pi \mu\right)}\,,\qquad\quad
C_2 = C_1 \, e^{-2i\pi \mu}
\,\frac{2\mu-1}{2\mu+1}\,\frac{
\Gamma\left(\tfrac{1}{2}-\mu\right)^2}{
\Gamma\left(\tfrac{1}{2}+\mu\right)^2}\,.
\ee
It is readily checked that the above linear combination yields a real
function\footnote{To check the $z\to 2$ asymptotics and reality of
  the solution, one must 
  choose the correct branch of the hypergeometic function
  which has a branch point at $z=2$ (when its argument diverges).}
 which is regular at $z=2$. Near the AdS boundary ($z\to 0$), we can
 then read
 off the coefficients in the asymptotic expansion \eqref{asymp}. After
 reinstating the radius $R$ of the boundary three sphere which was
 effectively set to $1$ in our conventions, we obtain the
 following analytic expressions for the mass $m$ and the condensate
 $\langle\tilde\psi_q\psi_q\rangle$ 
\\
\bea
\theta_{(0)} &=& 
-\Upsilon
\,\frac{8}{\pi}
\,\frac{\cos\,{\pi\mu R}\,}{4\mu^2 R^2-1}\,,
\\\nonumber\\\nonumber
\theta_{(2)}&=& - \Upsilon\,\frac{
\cos \pi\mu R}{
\pi
(4\mu^2R^2-1)}\,\times\,
\\\nonumber\\\nonumber
&\times&\,\left[3- (4\mu^2R^2-1)\left\{
\psi\left(\tfrac{1}{2}-\mu R\right)
+\psi\left(\tfrac{1}{2}+\mu R\right)+\ln 4+2\gamma_E-1\right\}\right]\,. 
\label{analytic}
\eea
Here $\psi(x)$ is the digamma function which has simple poles at $x=-n$, for 
all $n\in {\mathbb Z}$. Notice that in the expression for
$\theta_{(2)}$ above, these poles are precisely cancelled by the zeroes
of the cosine and $\theta_{(2)}$ is finite for all values of
$\mu$. 

Importantly, the set of linearized solutions, parametrized by $\Upsilon$,
yields massless hypermultiplets precisely when 
\be
\theta_{0} =0 \,\implies 2\mu R = 2\ell+1\,,
\qquad\ell=1,2,3,\ldots 
\ee
Therefore, non-constant massless embeddings first appear for 
half-integer values of the R-charge chemical potential. This is in
agreement with the numerical results from the complete DBI action,
which exhibit a more intricate structure including metastable and
unstable regions in the vicinity of the same critical values.
It is also easily checked that for these massless embeddings 
the chiral condensate
is non-zero:
\be
\langle\tilde\psi_q\psi_q\rangle = \,\Upsilon
\,\frac{\lambda N\,N_f}{2(2\pi)^4 R^3}\,(-1)^\ell\,.
\ee

To summarize, the large $\mu$ linear analysis has shown that
a continuous family of massless configurations, parametrized by $\Upsilon \leq
\sin^{-1}\left(1/2\mu\right)$ and carrying a finite R-charge density, emerges
 at the half-integer values of $\mu R$ indicated above. 
In contrast, the non-linear DBI equations show
a similar set of (unstable) 
solutions spread out in a small window in the vicinity of
these critical values. 

\begin{figure}[h]
\begin{center}
\epsfig{file=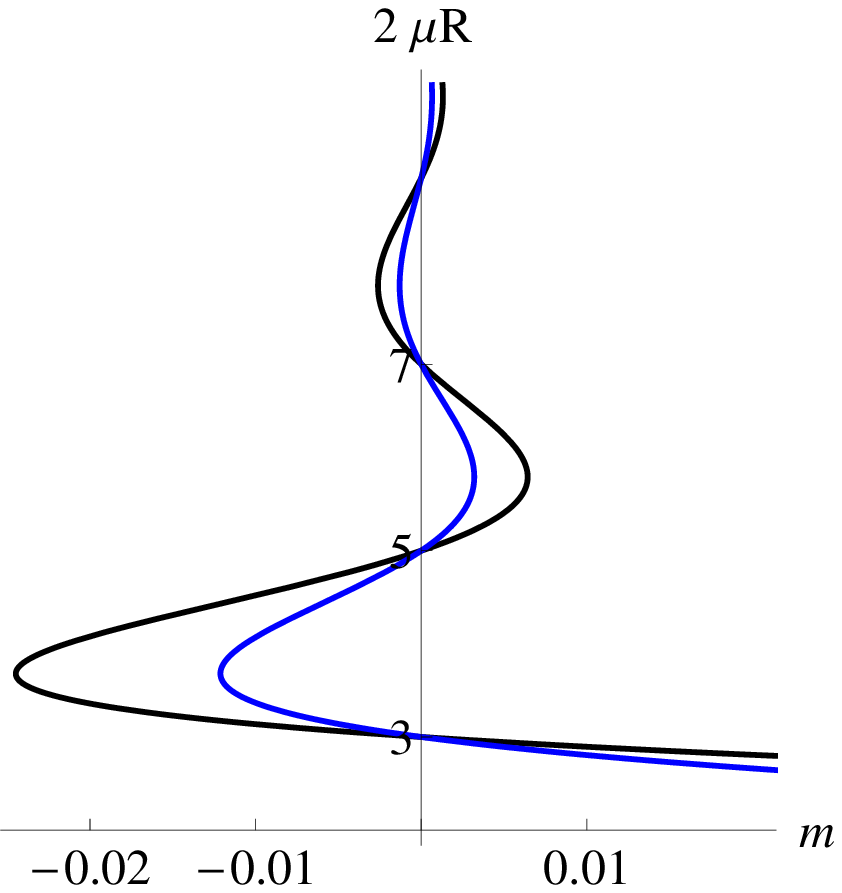, width =2.1in}
\hspace{0.5in}
\epsfig{file=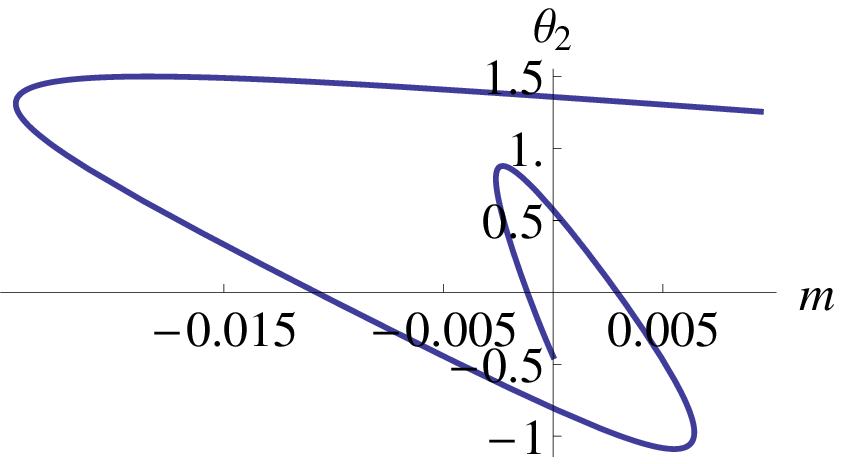, width =2.1in}
\end{center}
\caption{Plots of the expressions, obtained in the linearized
  approximation.
}
\label{exact}
\end{figure}

The analytical expressions for the mass and the chiral condensate in
Eq.\eqref{analytic}, 
reproduce quite closely the numerical results displayed in the
previous sections (see Fig.\eqref{exact}).

\subsection{Stability analysis: two-point function for 
$\tilde\psi_q\psi_q$}

The detailed numerical study of the free energies of the various
D7-brane embeddings, summarized in Fig.\eqref{saddle} suggests that the
values $2\mu R=3, 5, 7, \ldots$ are critical points where two saddle
points of the free energy merge. We expect that small fluctuations
around such points should cost no free energy at linear order, 
which is essentially what our
linearized analysis above reveals. If the chemical potential were to
be increased beyond these critical values, the system should become
thermodynamically unstable. The natural object to compute, to look 
for the source of instabilities in the system is the two-point
function of the operator dual to the slipping mode. In particular,
the location of the poles in the frequency-space correlator of
${\tilde\psi_q\psi_q}$ will reveal the presence of 
instabilities in the system. Actually, the linearized analysis above can
naturally be adapted to allow for a holographic evaluation of the two
point function of the operator dual to $\theta$. 

Since the dual field theory is formulated
on the three-sphere, it is natural to consider temporal 
correlations of the $s$-wave mode of ${\cal O}\equiv\tilde\psi_q\psi_q$:
\be
{\cal G}(t) = \int \int \frac{d^3\Omega}{\rm
  Vol(S^3)}\,\frac{d^3\Omega'}{\rm Vol(S^3)}\,  
\langle{\cal O}(t,\Omega){\cal O}(0,\Omega')\rangle\,.
\ee
We begin by expanding the DBI
action to quadratic order in $\theta$, allowing for the slipping angle
to be time dependent, so that $\theta=\theta(z,t)$. The induced metric
on the D7-brane is then
\bea
ds^2\Big|_{\rm D7}&=&
-\left(\frac{(1+\tfrac{1}{4}\,z^2)^2}{z^2}-4\mu^2\sin^2\theta -
\dot\theta^2\right)dt^2
+\left(z^2 \theta'(z)^2+ 1\right)\frac{dz^2}{z^2}+\\\nonumber 
&+& 2\dot\theta\theta' \,dt\,dz\,+\,\frac{(1-\tfrac{1}{4}\,z^2)^2}{ z^2}d\Omega_3^2+\cos^2\theta d\tilde
\Omega_3^2.
\eea
At quadratic order, the DBI action yields
\bea
&&S_{\rm DBI}\simeq \\\nonumber
&&2\pi^2\,{\cal N}_{\rm D7}\,\left[\int dt \,dz\,
\frac{(1 - \tfrac{1}{4}z^2)^3(1+\tfrac{1}{4}z^2)}{z^5} 
\left(\,z^2\,\theta^{\prime 2} -
  \frac{z^2}{(1+\tfrac{1}{4}z^2)^2}(\dot\theta^2+4\mu^2\theta^2) 
- 3\,\theta^2)\right)
\right]\,.
\eea
This is precisely the quadratic action for a charged scalar in global
$AdS_5$ with
$m^2_{\rm scalar}=-3$, coupled to a constant, background 
gauge potential $2\mu$. This is natural, given the interpretation of
$\mu$ as a chemical potential for a global $U(1)$ symmetry in the
boundary theory. The extra factor of 2 is the charge of the fermion
bilinear under the $U(1)$ R-symmetry in the boundary.
This type of system is also encountered in
holographic superconductor models, where there is typically 
a dynamical bulk Maxwell field
coupled to a charged scalar \cite{sean,hintro}, leading to a VEV
for the latter. The non-trivial embeddings with an R-charged
condensate appear to be of a similar nature albeit in global AdS
spacetime without a black hole.

Upon Fourier transforming to frequency space,
\be
\tilde \theta(z,\omega)=\int_{-\infty}^\infty dt\,e^{-i\omega t} \theta(z, t),
\ee
we note that at this order the effect of a non-zero 
temporal frequency is a simple shift,
\be
\mu^2 \to 4\mu^2+\omega^2,
\ee
and $\tilde \theta(z,\omega)$ satisfies the linearized equation of
motion Eq.\eqref{lineareq} after the above replacement. Therefore, 
 for the holographic calculation of the frequency space propagator we
can simply import the relevant results from the previous section and
make the replacement $\mu \to \tfrac{1}{2}\sqrt{4\mu^2+\omega^2}$.  
The near boundary expansion of $\tilde\theta(z,\omega)$,
\be
\tilde \theta(z,\omega)\big|_{z\to 0}\simeq 
\Theta_0(\omega)\,z + \Theta_2(\omega)\, z^3 +
\tfrac{1}{2
}\,(1-4\mu^2)\Theta_0(\omega)\, z^3\,
\ln z + \ldots.
\ee
can be determined precisely from 
our linearized analysis above. In particular, we already know that
regularity at the origin of AdS space leads to a relation between
$\Theta_2$ and $\Theta_0$, from Eq.\eqref{analytic}:
\bea
\Theta_2(\omega) \,&=&\,\Theta_0(\omega)\times\\\nonumber
&\times&\left[3- (4\mu^2+\omega^2-1)
\left\{\psi\left(\tfrac{1-\sqrt{4\mu^2+\omega^2}}{2}\right)+
\psi\left(\tfrac{1+\sqrt{4\mu^2+\omega^2}}{2}\right)+\ln 4+2\gamma_E-1\right\}\right]\,.
\eea
Substituting this into the linearized action above, we evaluate the
action induced on the boundary, 
\be
S_{\rm bdry} = 2\pi^2\,{\cal N}_{\rm D7}\,\int_{-\infty}^\infty 
\frac{d\omega}{2\pi}\,\frac{1}{\epsilon^3}\,\tilde\theta(\epsilon,
\omega) \, \tilde\theta'(\epsilon, -\omega)\,\big|_{\epsilon\to 0}+\ldots\,.
\ee
where we have not explicitly written out the counterterms.
Differentiating twice with respect to $\Theta_0(\omega)$ which acts as
the source for $\tilde \psi_q\psi_q (\omega)$, results in the frequency space
correlator. We write the result after re-introducing the radius $R$ of the
boundary three-sphere and ignoring ``contact'' terms which are regular
in $\omega$:
\be
\tilde {\cal G}(\omega) = \,{\cal N}_{\rm D7}
(4\mu^2+ \omega^2- R^{-2})
\left[\psi\left(\tfrac{1}{2}- \tfrac{1}{2}R
\sqrt{4\mu^2+\omega^2}\right)+
\psi\left(\tfrac{1}{2}+ \tfrac{1}{2}R\sqrt{4\mu^2+\omega^2}\right)\right]\,.
\ee
Despite the appearance of square roots in this expression, the
function only has isolated simple poles at 
\be
\omega = \pm \sqrt{R^{-2}(2\ell+1)^2-4\mu^2}\,,\qquad \ell=1,2,\ldots.
\ee
\begin{figure}[h]
\begin{center}
\epsfig{file=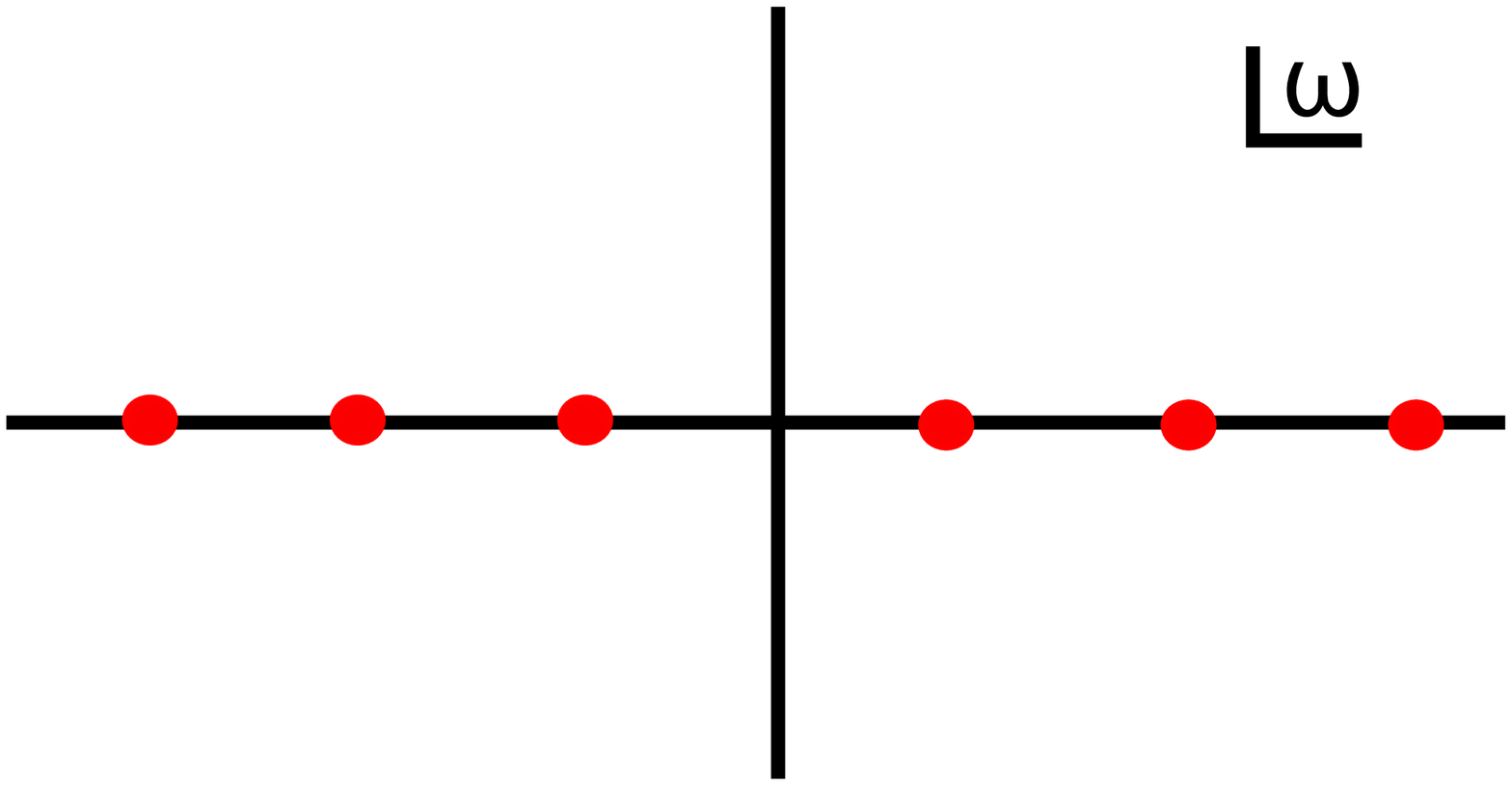, width =2.1in}
\hspace{0.5in}
\epsfig{file=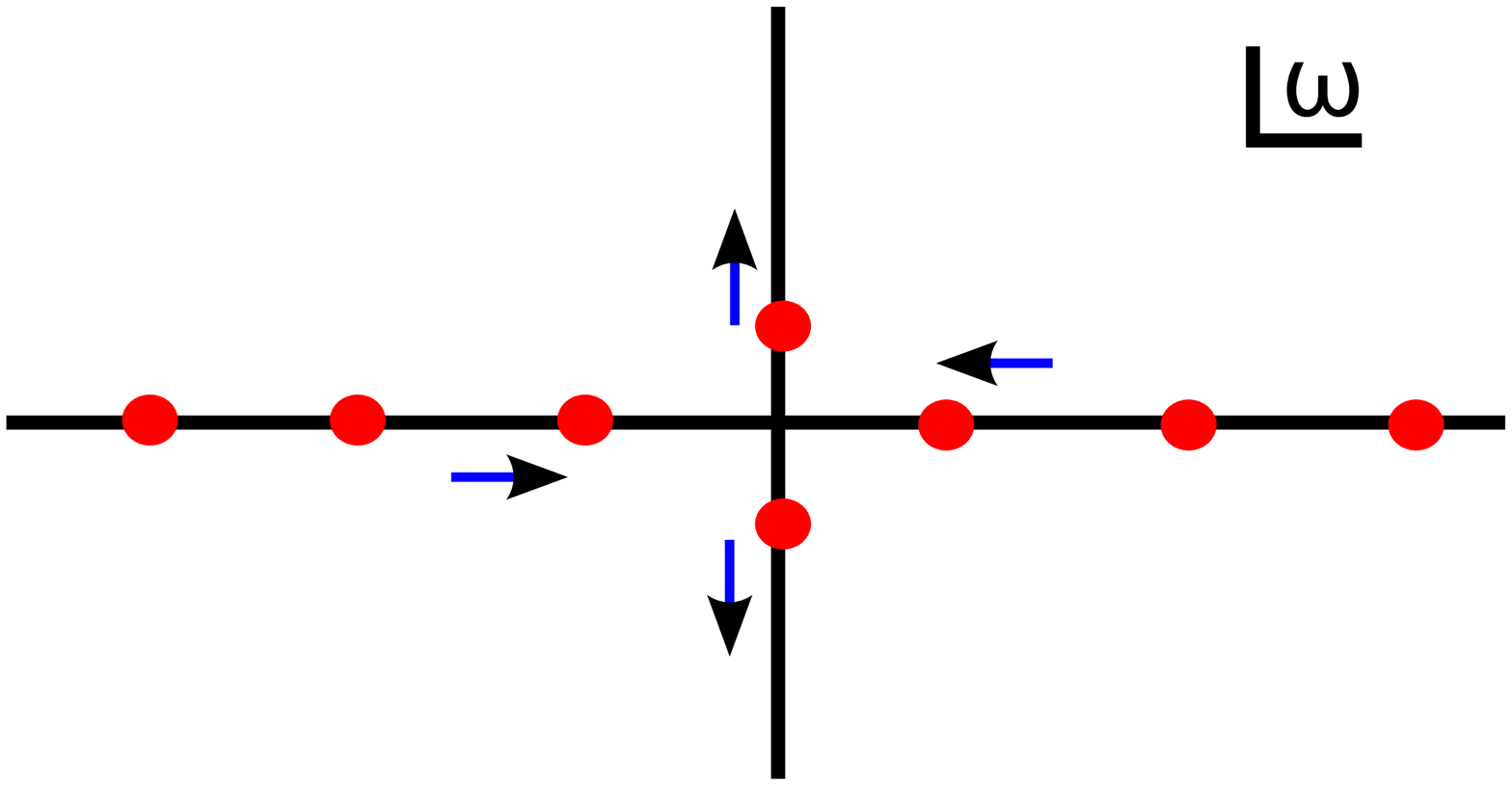, width =2.1in}
\end{center}
\caption{ Simple poles of $\tilde{\cal G}(\omega)$ on the real axis, migrate
  to the imaginary axis as $\mu$ is increased past $\frac{3}{2} R^{-1}$.
}
\label{poles}
\end{figure}
For $\mu >\frac{3}{2} R^{-1}$, some of the poles move to the upper
half plane, along the imaginary axis, reflecting an instability.

The $\mu=0$ functional 
form of the correlator coincides with the expression deduced from the
weakly coupled field theory\footnote{In recent work \cite{erdfilev} 
  the meson spectrum of probe branes in global AdS was obtained. 
 The poles in the
  correlator we find here correspond to $l=\tilde l=0$ scalar states in
that work.} in Eq.\eqref{weakgf}. 
However, the normalizations differ between weak and strong
couplings. Therefore the correlators do get renormalized even in the absence of
a chemical potential.

In the strongly interacting system, the chemical potential has the
effect of shifting the locations of the poles of the Green's function. With
$\mu\neq 0$, there is a qualitative change in the physics in going to
strong coupling. The discrete excitations at strong coupling consist of
mesons/fermion bilinears made from the spherical harmonic states of
elementary fermions. These composite scalars carry twice the R-charge of the
fermions, and the poles of the correlator show that they 
possess twice the energy. Importantly, the R-charge chemical potential
affects the masses of these states in the same way that it would alter
the mass of a free complex scalar. A large enough chemical potential
(bigger than the mass to charge ratio for these states) 
leads to a negative mass squared and triggers an instability favouring a
condensate. 

We expect that the instability persists when a flat space limit is
taken, by formally letting $R$ become arbitrarily large, so that
\be
\tilde{\cal G}(\omega)\to {\cal N}_{\rm D7}\,(\omega^2+4\mu^2)\,
\ln\left(\omega^2+4\mu^2\right),
\ee
which has non-analyticities in the upper half plane.

It is interesting to contrast the above picture with the physics of
the free theory
where the positions of the poles are unaltered by a
non-vanishing chemical potential. The primary effect there is the Pauli
blocking of low lying states, which in turn results in an imaginary
part for the retarded Green's function in flat space beyond a
kinematic threshold $|\omega|> 2\mu$ (Eq.\eqref{flatweak}).

\section{D5 and D3$^\prime$ flavour branes in global $AdS_5$}
Most aspects of the analysis we performed for probe D7-branes
above,  carry over to probe D5 and D3-branes whose excitations in $AdS_5$
describe 2+1 and 1+1-dimensional defect CFTs 
coupled to ${\cal N}=4$
SYM at strong coupling \cite{DeWolfe:2001pq, Constable:2002xt,
  Karch:2009ph}. 
In the global AdS context, the defect CFT's are
formulated  
on a spatial $S^2 \subset S^3$ and $S^1\subset S^3$.   
The eigenvalues of the Dirac operator on $S^2$ take integer values
\cite{camporesi}, so that
\be
\vep_\ell\,(S^2)= \ell\,R^{-1}\,,\qquad\ell=1,2,\ldots
\ee 
For the 1+1 dimensional case, fermion energies on the $S^1\subset S^3$
are half-integer valued
\be
\vep_\ell\,(S^1)= (\ell-\tfrac{1}{2})\,R^{-1}\,,\qquad\ell=1,2,\ldots
\ee 
Interestingly, the probe D-branes exhibit the same kind of
instabilities we saw above for the D7-branes, but now at values of the
chemical potential set by the fermion energy levels on $S^2$ and
$S^1$. We will not repeat the detailed numerical study of these
examples, since their basic features turn out to be quite similar to
the probe D7-brane. Instead, we will briefly explain the analytical
results for the lower dimensional examples and point out features
which are distinct from the four dimensional case.

\subsection{D5-brane probes}
The 2+1 dimensional CFT localized on the intersection of a large-$N$
number of D3-branes with $N_f$ D5-branes has eight supercharges and
an $SU(2)_H\times SU(2)_V$ R-symmetry. The defect theory has
fundamental hypermultiplets, each containing a fermion doublet
transforming under $SU(2)_V$. We wish to introduce a chemical
potential for the $U(1)$ subgroup of this $SU(2)_V$
R-symmetry under which the fundamental representation 
fermions are charged. As shown in \cite{DeWolfe:2001pq} the scalars in
the fundamental hypermultimplets are uncharged under this. In the 
near horizon limit of this system at large-$N$, probe D5-branes in
(the Poincare' patch of) $AdS_5\times S^5$ can therefore be chosen to
spin along the internal directions so that R-charge is
concentrated in the fermionic sector of the fundamental matter
fields.

The same setup, when carried over to global AdS spacetime, involves
flavour D5-branes with world-volume global $AdS_4\times S^2$. The
$S^2 \subset S^5$ wrapped by the branes 
can be parametrized in terms of the angular coordinate $\theta$ as,
\be
ds^2_{S^5}= d\theta^2+ \sin^2\theta\,d\tilde\Omega_2^2 + 
\cos^2\theta\,d\Omega^2_2\,.
\ee
$\theta$ plays the role of the ``slipping'' angle and the
D5-branes can be chosen to spin along the azimuthal angle $\phi$ inside
the transverse two-sphere $\tilde S^2$. The induced metric for the spinning
D5-brane with world-volume $AdS_4\times S^2$ is
\bea
&&ds^2\Big|_{\rm D5}=\\\nonumber
&& -\left(\frac{(1+\tfrac{1}{4}z^2)^2}{z^2}-4\mu^2\sin^2\theta\right)dt^2
+\left(z^2 \theta'(z)^2+ 1\right)\frac{dz^2}{z^2}+ 
\frac{(1-\tfrac{1}{4}z^2)^2}{z^2}d\Omega_2^2+\cos^2\theta d\Omega_2^2.
\eea
\subsubsection{Linearized results:}
The D5-brane probe action, in terms of embedding coordinates
$\{\xi_i\}$, is  
\be 
S_{\rm D5}= N_f\, T_{\rm D5}\,\int d^5\xi\,
\sqrt{-{\rm det}{}^*g}\,,\qquad 
{\cal N}_{\rm D5} \equiv \,N_f\,T_{\rm D3}\,4\pi\,=\,\frac{\sqrt\lambda}{4\pi^3}\,N_f\,N,
\ee
and the resulting non-linear equations of motion can be numerically
solved. Similarly to the D7-brane case, we look for embeddings of the
slipping mode $\theta(z)$ in the linearized approximation. This
 mode maps to a dimension two R-charged operator in the
boundary 3D CFT, namely the quark (fermion) bilinear $\tilde \psi_q\psi_q$.
The
solutions of interest are those that get to the origin of $AdS$. At
quadratic order the spinning D5-brane action describes a charged
scalar in $AdS_4$, with $m^2_{\rm scalar}=-2$, in the presence of a
constant background gauge potential $2\mu$. The linear equation of
motion,
\be
\theta''(z) -2\frac{16+ 4z^2+ 2 z^4}{z(16-z^4)}\,\theta'(z)+
2\frac{16+z^4+8 z^2(1+4\mu^2)}{z^2(4+z^2)^2}\, \theta(z)=0,
\ee
subject to the boundary conditions,
\be
\theta'(z)\big|_{z=2}=0\,,\qquad \theta(z)\big|_{z=2}= \Upsilon\,,
\ee
 is solved by 
\be
\theta(z)\, =\, 
4\,\Upsilon\, \frac{z}{\mu}\frac{\sin \left(4\mu\,\tan^{-1}\,\tfrac{z}{2} -
\frac{\pi\mu}{2}\right)}{(z^2-4)}\,.
\ee  
The asymptotic expansion near the boundary then yields the mass and
the fermion bilinear condensate,
\bea
&&\theta(z)\simeq \theta_{(0)}\, z + \theta_{2}\,
z^2+\ldots\\\nonumber\\
&&\theta_{(0)}\,=\, \Upsilon\, \frac{\sin\,(\pi\mu R)}{\mu R}\,,\qquad
\theta_{(2)}\,=\, - \Upsilon \cos\,(\pi \mu R) \label{masscond}
\eea
Therefore the linearized embeddings are massless only 
when,
\be
\mu R \,=\, \ell R^{-1}\,\qquad\qquad\ell=1,2,3,\ldots
\ee
As before, these are critical values of the chemical potential beyond
which  corresponding fermion bilinears will become unstable, as we
show below.
For these finite density embeddings the chiral condensate is non-zero,
\be
\langle\tilde\psi_q\psi_q\rangle=
\,-{\cal N}_{\rm D5}\, (-1)^\ell\qquad\qquad\ell \in {\mathbb Z}.
\ee
\subsubsection{Stability of $\theta=0$ embedding:}

The calculation of the $s$-wave 2-point function in frequency space, 
for the fermion bilinear
${\cal O} = \tilde\psi_q\psi_q$ proceeds exactly as in the 3+1 dimensional
theory. The result is obtained by taking the ratio of the normalizable
and non-normalizable modes in \eqref{masscond}, followed by the
replacement $\mu \to \tfrac{1}{2} \sqrt{4\mu^2+\omega^2}$. We find,
\be
\tilde{\cal G}(\omega) \,=\, {\cal N}_{\rm D5}\,\sqrt{4\mu^2+\omega^2}\,
\cot(\tfrac{\pi}{2} R\sqrt{4\mu^2+\omega^2})
\ee
which has simple poles at 
\be
\omega = \pm \sqrt{(2\ell)^2 \,R^{-2}- 4\mu^2}\,,\qquad \ell=1,2,\ldots.
\ee
Therefore the onset of instabilities occurs when $\mu R > 1$, when the
poles move to the upper half plane. Once again the
chemical potential acts as a negative mass squared for the
scalar excitations at strong coupling
\subsection{D3$^\prime$ probe branes}
The theory on the intersection of the D3-D3$^\prime$ system was
studied in detail in \cite{Constable:2002xt}. In flat space, the
D-brane setup involves two stacks with $N$ and $N_f$ D3-branes,
spanning the $x^0,x^1,x^2, x^3$  and $x^0,x^1,x^4,x^5$ 
directions respectively. Each stack yields an ${\cal N}=4$ vector
multiplet in four dimensions, coupled to $SU(N)\times SU(N_f)$ 
bifundamental degrees of freedom localized at the intersection of the
two stacks. 

In the 't Hooft large-$N$ limit with $N_f$ fixed, the $SU(N_f)$ vector
multiplet decouples since its effective coupling $\lambda_f \equiv\lambda
N_f/N$ is vanishing in the large-$N$ limit. Therefore, one
obtains ${\cal N}=4$ SUSY Yang-Mills theory in four dimensions 
coupled to $N_f$ two dimensional 
${\cal N}=(4,4)$ fundamental hypermultiplets localized at the
intersection of the two sets of D3-branes. In this limit, and at
strong 't Hooft coupling, the embedding of the 
$N_f$ probe D3$^\prime$-branes in the $AdS_5\times S^5$ geometry
induces an $AdS_3\times S^1$ metric on the probes.  The $S^1\subset
S^5$, wrapped by the D3$^\prime$ s can be specified as usual by the
slipping angle $\theta$
\be
d\Omega_5^2 = d\theta^2 + \sin^2\theta\, d\Omega_3^2 + \cos^2\theta\,d\Omega_1^2\,.
\ee 
The global isometry of the coordinates transverse to all the D3-branes
is $SO(4)\simeq SU(2)_L \times SU(2)_R$. Under this isometry, the
fermions in the fundamental $(4,4)$ hypermultiplets transform in the 
$(\frac{1}{2},0)$ and $(0,\frac{1}{2})$ representations
respectively. Their scalar superpartners are uncharged under this.
The fermion bilinear, which we will continue to denote
schematically as $\tilde \psi_q\psi_q$, dual to the slipping mode,
transforms in the $(\frac{1}{2},\frac{1}{2})$ representation of
$SU(2)_L\times SU(2)_R$. It has scaling dimension $\Delta=1$,
consistent with a fermion bilinear in two dimensions, and is a
$\tfrac{1}{2}$-BPS operator. 

The above system in global AdS spacetime is dual to the 1+1
dimensional defect CFT formulated on an equatorial circle of the
boundary three-sphere.
Imparting an R-charge to the flavour fermions by rotating the 
D3$^\prime$'s along an angular coordinate $\phi$ in the $S^3$ with angular
velocity $2\mu$, the induced metric on the D3$^\prime$ becomes
\bea
&&ds^2\Big|_{{\rm D3}^\prime}=\\\nonumber
&& -\left(\frac{(1+\tfrac{1}{2}z^2)^2}{z^2}-4\mu^2\sin^2\theta\right)dt^2
+\left(z^2 \theta'(z)^2+ 1\right)\frac{dz^2}{z^2}+ 
\frac{(1-\tfrac{1}{2}z^2)^2}{z^2}d\Omega_1^2+\cos^2\theta d\Omega_1^2.
\eea

This yields the probe D-brane action
\be
S_{{\rm D3}^\prime}= \,N_f \,T_{\rm D3}\int\,d^4\xi\,\sqrt{-{\rm
    det}{}^*g}\,,\qquad T_{{\rm D3}^\prime}=\frac{N}{2\pi^2}\,.
\label{dbid3}
\ee
The slipping mode, at linearized order, is a scalar with $m^2_{\rm
  scalar}=-1$ in $AdS_3$. It therefore saturates the relevant
Breitenlohner-Freedman bound with boundary asymptotics,
\be
\theta(z)\to \theta_{(0)}\,z\,\ln z + \theta_{(2)}\,z+\ldots
\ee
Unlike the higher dimensional cases, the term linear in $z$ is now
normalizable, whilst the non-normalizable mode is the coefficient of
the asymptotic logarithmic term. The latter plays the role of the
hypermultiplet mass. When $\mu \gg 1$, it is consistent to linearize
the DBI equation of motion, 
\be
\theta''(z)-\frac{(16+3z^4)}{z(16-z^4)}\,\theta'(z)- 
\frac{16+8z^2(1+8\mu^2)+z^4}{z^2(4+z^2)^2}\,\theta(z)=0\,.  
\ee
The solutions to this equation, which fall in the category of
``AdS-filling'' embeddings, can be written in terms of
hypergeometric functions satisfying the boundary conditions,
\be
\theta(z)\big|_{z=2}\,=\,\Upsilon\,\qquad \theta'(z)\big|_{z=2}\,=\,0\,.
\ee
The corresponding solution is 
\bea
&&\theta(z) \,=\,-\Upsilon\,\frac{i}{2}\,\frac{\sqrt{y-1}}{\sin
  \pi\mu}\,\left( \,
\frac{e^{i\pi\mu}\,y^{\mu}\,\Gamma\left(\tfrac{1}{2}+\mu\right)}
{\Gamma\left(\tfrac{1}{2}-\mu\right)\Gamma\left(1+2\mu\right)}
\,\,\,{}_2F_1\left(\tfrac{1}{2}+\mu,\tfrac{1}{2}+\mu\,;1+2\mu\,;y\right)
\right.\\\nonumber
&&\left.\qquad\hspace{3.3in}- (\mu \to -\mu)\right)\\\nonumber
&& y \equiv \frac{(1+\tfrac{1}{4}z^2)^2}{(1-\tfrac{1}{4}z^2)^2}\,.
\eea
From the asymptotic expansion of this function, we read off the
fermion mass and condensate
\bea
&&\theta_{(0)}\,=\,- \frac{2}{\pi}\,\Upsilon\,\cos(\pi\mu
R)\,,\\\nonumber
&&\theta_{(2)}\,=\,-\frac{1}{\pi}\Upsilon\,\cos(\pi\mu R)\left[
\psi\left(\tfrac{1}{2}+\mu R\right)+
\psi\left(\tfrac{1}{2}-\mu R\right)+2\gamma_E+\ln 4\right]\,.
\eea
As for probe D5 and D7-branes, the VEV and the mass satisfy a relation
\be
\theta_{(2)} \,=\,\tfrac{1}{2}\,\theta_{(0)}\,\left[
\psi\left(\tfrac{1}{2}+\mu R\right)+
\psi\left(\tfrac{1}{2}-\mu R\right)+2\gamma_E+\ln 4\right]\,.
\label{relation}
\ee
Therefore the embeddings yield zero mass hypermultiplets when 
\be
2\mu R = 2\ell -1\,,\qquad \ell=1,2,3,\ldots,
\ee
which is in agreement with what we would expect for fermion bilinears 
on $S^1$. To correctly calculate both one- and two-point functions of
the operator $\tilde\psi_q\psi_q$, it is useful to recall the
holographic renormalization procedure for probe branes
\cite{karchskenderis} applied to the D3-D3$^\prime$ system. This is
particularly relevant in this case
due to putative logarithmic divergences in 1+1 dimensions. The
regularized DBI action is obtained by performing a subtraction at 
$z=\epsilon \ll 1$,
\be
S^{\rm reg}_{{\rm D3}^\prime} \,=\,4\pi^2\,N_f\,T_{\rm
  D3^\prime}\,\int dt\,\left(\int_\epsilon^2 dz\,\sqrt{-{\rm det}{}^*g}
-\frac{1}{2\epsilon^2}+\frac{1}{2\epsilon^2}\,\theta(\epsilon,t)^2
\,\left(1+\tfrac{1}{\ln \epsilon}\right)
\right)
\ee
Using this, it follows that 
\be
\langle\tilde\psi_q\psi_q \rangle \,=\, \lim_{\epsilon\to 0}\,
\frac{\ln\epsilon}{\epsilon}\,\frac{1}{\sqrt {-\gamma}}\,
\frac{\delta S^{\rm reg}_{\rm D3^\prime}}{\delta \theta(\epsilon,
  t)}\,=\,
\frac{N\,N_f}{2\pi R}\,\theta_{(2)},
\ee
where we have restored explicit dependence on the radius of the
boundary circle.  
Therefore for the massless, finite density embeddings in the vicinity
of  $ 2\mu R =
2\ell-1$ in the linearized
approximation, the chiral condensate is
\be 
\langle\tilde\psi_q\psi_q\rangle\big|_{m=0} \,
=\, \Upsilon\,\frac{N\,N_f}{2\pi R}\,
(-1)^\ell\,,\qquad\ell\in{\mathbb Z}
\ee

\subsubsection{Two-point function around $\theta=0$:}
It is now straightforward to obtain 
the two-point function of the fluctuations of 
${\cal O}=\tilde\psi_q\psi_q$, about the trivial
zero-density embedding in frequency
space. A time dependent fluctuation $\theta(z,t)$ about the $\theta=0$
solution, can be expanded near $z=0$ as (in frequency space),
\be
\tilde\theta(z,\omega)
\,=\,\Theta_0(\omega)\,z\ln z +\Theta_2(\omega)\, z+\ldots
\ee
The VEV of ${\cal O}(\omega)$ and the source $\Theta_0(\omega)$ obey
the relation \eqref{relation} with the replacement $\mu \to
\tfrac{1}{2}\sqrt{4\mu^2+\omega^2}$. The Green's function for ${\cal O}$ is then
given by 
\bea
&&\langle{\cal O}(\omega){\cal O}(-\omega)\rangle\,=
-\frac{1}{2}\,\frac{\delta\langle{\cal O}(\omega)\rangle}{\delta \Theta_0(-\omega)}\\\nonumber\\\nonumber
&&= 
- \frac{N\,N_f}{4\pi R}\,\left[
\psi\left(\tfrac{1}{2}+\tfrac{1}{2}R\sqrt{4\mu^2+\omega^2} \right)+
\psi\left(\tfrac{1}{2}-
\tfrac{1}{2}R\sqrt{4\mu^2+\omega^2} \right)+2\gamma_E+\ln 4\right]\,.
\\\nonumber
\eea
As was noted in \cite{Constable:2002xt}, the Green's functions and
indeed, the DBI action associated to probe D3-branes \eqref{dbid3} 
is independent of the 'tHooft coupling $\lambda$, suggesting a
non-renormalization theorem. However, this could only possibly be true
at $\mu=0$, in the supersymmetric theory. In the situation with a
chemical potential we see that the locations 
of poles in the correlator depend non-trivially on $\mu$. In particular,
as in the higher dimensional examples, the poles of the Green's function
shift to 
\be
\omega = \pm \sqrt{(2\ell -1)^2R^{-2}-4\mu^2}\,,\qquad\ell =1,2,3,\ldots
\ee
which imply instabilities when $\mu R > \tfrac{1}{2}$.

\section{Discussion}
We have seen that spinning Dp-branes in global $AdS_5\times S^5$, dual
to strongly coupled CFT's on spheres with a chemical potential for
fermion flavours alone, exhibit instabilities (preceded by
metastabilities). 
It appears that in all cases the zero density state becomes metastable or
unstable when the chemical potential approaches integer or
half-integer values, depending on the number of spatial dimensions in
the boundary CFT. At each of these points, the systems would prefer to
be in a configuration with finite R-charge density and with a
homogeneous condensate made from harmonics of the elementary fermions
on the spheres. 

However, the probe DBI analysis leaves the central question
unanswered, namely, what is the end-point of the unstable directions. 
The answer to this is not clear from the above computations,
since the fully non-linear DBI solutions show a runaway effective
potential. One plausible explanation could be that determining the
true ground state might require new active ``light'' modes on the probe
brane. This could happen, for example, if the ground state were
inhomogeneous, as is the case in certain lower dimensional models
\cite{dunne}

However, a possible answer may lie in the differing, and somewhat
complementary, physical pictures that emerge 
at arbitrarily weak or zero coupling
and strong coupling. In the free theory, the relevant degrees of
freedom are the elementary fermions (and their spherical harmonics),
whereas in the strongly interacting AdS/CFT dual, the D-brane 
``slipping mode'' is the natural degree of freedom. This is a
composite scalar - a fermion bilinear. A large enough R-symmetry
chemical potential, effectively 
induces a negative mass squared for this composite
scalar. This situation is reminiscent of the Gross-Neveu model 
\cite{Gross:1974jv} where
the elementary fermions are traded for a composite scalar and an
effective potential for this composite scalar is subsequently induced,
leading to chiral symmetry breaking. In that case, the full effective
potential requires resumming fermion (flavour) loops. It is
conceivable that in the present context similar, subleading
corrections involving fermion loops need to be incorporated into the
picture. This would also potentially mean inclusion of 
back-reaction effects of flavour branes on the background
geometry. The possibility remains, of approaching this question from
the weakly coupled boundary theory where effective four-fermi
interactions can be induced via the Yukawa couplings of the flavour
fermions to the adjoint modes of the ${\cal N}=4$ theory on the sphere.
\\

{\bf Acknowledgements:} I would like to thank Gert Aarts, Paolo
Benincasa, Simon Hands, Tim Hollowood, Andreas Karch, 
Andy O' Bannon, Jim Rafferty 
and Larry Yaffe for discussions and useful comments. 
Special thanks to Andy O' Bannon and Paolo Benincasa 
for insightful comments and suggestions on early drafts of the
manuscript. I also acknowledge the stimulating atmosphere provided by
the Erwin Schr\"odinger Institute, during the
EMMI Workshop on ``Hot Matter: Quasiparticles or Quasinormal modes'',
August 2010, where some of this study was initiated.
This work is supported by STFC Rolling Grant ST/G0005006/1.

\newpage
\startappendix
\Appendix{Weak coupling Gross-Witten transitions}
\label{appa}
In \cite{Hands:2010zp}, it was shown that large-$N$ Yang-Mills theory
on $S^3$ with $N_f$ fermions in the fundamental representation ($N_f/N$ fixed),
undergoes an infinite sequence of Gross-Witten transitions as a
function of a baryon number chemical potential, at a fixed low
temperature. In that case, the Euclidean path integral was shown to be
dominated by a {\it complexified}
saddle point configuration of eigenvalues of the 
Polyakov-loop at large-$N$. The complex nature of the saddle point was
due to the fact that the Euclidean action with a baryon number
chemical potential, is not Hermitian - the source of the ``sign problem''.

Below, we will repeat the essential aspects of this computation for
the case where a chemical potential is introduced for 
an axial $U(1)$ symmetry for the fermions. We take $N_f$ flavours of
massless Dirac fermions 
\bea
\psi_D^i\,=\,\left(\begin{array}{ccc}
\psi_q^i\\\\
\tilde\psi_q^{i\,\dagger }
\end{array}
\right)\,\qquad i=1,2,\ldots N_f.
\label{dirac}
\eea
In Lorentzian signature, a chemical potential $\mu$ for the axial $U(1)$
symmetry, is introduced by 
\be
{\cal L} \to {\cal L} +
\mu\,\sum_{i=1}^{N_f}\,\bar\psi_{D\,i}\,\gamma^0\gamma_5\psi_{D}^i +\ldots,
\ee
where the ``$\ldots$'' represent terms due to any additional matter fields
in the theory charged under the $U(1)$.
To make sense of this we need to assume that the symmetry is anomaly
free, which can be achieved in supersymmetric theories with an
appropriately chosen matter content (e.g. the matter content of ${\cal N}=2$
supersymmetric gauge theory with $SU(N)$ gauge group and $N_f=2N$
hypermultiplets). The main feature we want to illustrate is
largely independent of these details.

At finite temperature $T$, going to Euclidean signature, the theory is
formulated on $S^3\times S^1$. The circumference of the $S^1$ is
$\beta=1/T$\,. Allowing for a Wilson loop around the thermal circle,
we can choose it to be diagonal and homogeneous on $S^3$ 
\be
\frac{1}{{\rm Vol}(S^3)}\,\oint_{S^1\times S^3} A_0\, = \,{\rm
  diag}\,(\theta_1,\theta_2,\ldots\theta_N)\,,\qquad
\sum_{i=1}^{N}\theta_i \,=\, 0\,{\rm mod}\, 2\pi\,.
\label{eval}
\ee
Fermions do not have any zero modes on $S^3$ and at finite
temperature. Assuming that any other matter fields in the theory
do not have zero modes, we can integrate out all the Kaluza-Klein
harmonics on $S^3$ to obtain an effective unitary matrix model for the Polyakov
loop $U\equiv \exp{i\oint_{S^1} A_0}$. In the eigenvalue basis,
\eqref{eval} the effective action for the $\theta_i$ takes the form
\cite{sbh, Hollowood:2008gp, Yamada:2006rx, Hands:2010zp}
\bea
&&S_{\rm eff}[\theta_i]
\,=\,\sum_{i,j\,=1}^N\,-\,\tfrac{1}{2}\,
\ln\,\sin^2\left(\tfrac{\theta_i-\theta_j}{2}\right)
- \frac{N_f}{2}\,\sum_{j=1}^N\,d_\ell\,\ln \left(1+ e^{i\theta_j}\,
e^{-\beta(\vep_\ell-\mu)}\right)\\\nonumber
&& -\frac{N_f}{2}\,\sum_{j=1}^N\,d_\ell\,\ln \left(1+ e^{-i\theta_j}\,
e^{-\beta(\vep_\ell-\mu)}\right) \, - \,(\mu \to -\mu) \,+\,\ldots
\eea
The first term is the Van der Monde pairwise repulsive potential
between the eigenvalues, while the ``$\ldots$'' represent the
contributions from Kaluza-Klein harmonics of all other degrees of freedom
in the theory. The terms explicitly shown arise from integrating out the fermion
spherical harmonics, transforming in the fundamental representation of
$SU(N)$. The energies of the fermion modes and their degeneracies on
$S^3$ are
\be
\vep_\ell = \ell+\tfrac{1}{2}\,\qquad d_\ell=2\ell(\ell+1)\,,\qquad
\ell=0,1,2\ldots
\ee
We note that the effective action is real, in contrast
to the situation with baryonic chemical potential
\cite{Hollowood:2008gp}; nevertheless, the physics is quite similar to
that case. 

The physics of interest occurs at low $T$, when $\mu$ approaches
$\vep_\ell$. When $\mu \simeq \vep_\ell$, it suffices to
focus attention solely on the mode with that energy. In this regime
it is useful to introduce the fugacity
\bea
\zeta_\ell\,\equiv \, e^{-(\vep_\ell-\mu)/T}\,\qquad &&\zeta_\ell\, \ll\,1
\quad{\rm for}\qquad\mu\lesssim \vep_\ell\,,\\\nonumber
&&\zeta_\ell\, \gg\,1
\quad{\rm for}\qquad\mu\gtrsim \vep_\ell\,,
\eea
In the large-$N$ limit with $N_f\to \infty$ and $N_f/N$ fixed, the
saddle point equation for the effective action with
$\mu\simeq \vep_\ell$, is
\be
\sum_{j=1}^N
\,\frac{1}{N}\,\cot\left(\tfrac{\theta_i-\theta_j}{2}\right)\,\simeq\,
\frac{N_f}{N}\,d_\ell\,\frac{\zeta_\ell\sin\theta}{\zeta_\ell^2+2\zeta_\ell\cos\theta+1}\,+
\,{\rm exponentially \,\,small }
\ee
When $\mu < \vep_\ell$ ,  $\zeta_\ell \ll 1$ and the saddle point
equation becomes 
\be
\int_{-\pi}^\pi\,d\theta'\,
\rho(\theta')\,\cot\left(\tfrac{\theta-\theta'}{2}\right)\,
\simeq \, d_\ell\,\frac{N_f}{N}\, {\zeta_\ell}\,\sin\theta\,,\qquad
\zeta_\ell < 1.
\label{lowmu}
\ee
while for $\mu > \vep_\ell$
\be
\int_{-\pi}^\pi\,d\theta'\,
\rho(\theta')\,\cot\left(\tfrac{\theta-\theta'}{2}\right)\,
\simeq \, d_\ell\,\frac{N_f}{N}\, \frac{1}{\zeta_\ell}\,\sin\theta\,,\qquad
\zeta_\ell > 1.
\label{himu}
\ee
Here we have introduced the eigenvalue density $\rho(\theta)$ on the
circle, normalized so that $\int_{-\pi}^\pi d\theta\,\rho(\theta)
=1$. These are precisely the equations leading to the Gross-Witten
third order transition, \cite{Gross:1980he}. 

This analysis, (from the results of \cite{Gross:1980he}) shows that 
for both $\mu < \vep_\ell$, ($\zeta_\ell\ll 1$) and $\mu>\vep_\ell$, 
($\zeta_\ell\gg1$), the distribution function $\rho(\theta)$ should be 
``ungapped''. 

As $\mu$ is increased from low values towards $\vep_\ell$, we expect a
``gapping'' transition of third-order when
\be
\frac{N_f}{N}\,d_\ell\, {\zeta_\ell} \simeq 1\,.
\ee
which corresponds to 
\be
\mu\, =\,\mu^*_-\,\equiv\, \vep_\ell - T \ln
\left(\frac{N_f}{N}\,d_\ell\right)\,,
\ee
for small enough $T$. Therefore for $\mu > \mu^*_-$, the theory enters
a gapped phase for the eigenvalues of the Polyakov loop.

As $\mu$ is increased further past $\vep_\ell$, the effective
potential becomes of the form \eqref{himu} and now the theory
undergoes yet another transition from the gapped to an ungapped phase
at 
\bea
&&\frac{N_f}{N}\,d_\ell\, \frac{1}{\zeta_\ell} \simeq 1\,\\\nonumber
\implies &&\mu\, =\,\mu^*_+\,\equiv\, \vep_\ell + T \ln
\left(\frac{N_f}{N}\,d_\ell\right)\,.
\eea
Therefore, we learn that the free theory on $S^3$ at very low
temperatures, experiences a pair of Gross-Witten transitions as a
function of the chemical potential $\mu$, when the chemical potential
is in the vicinity of an energy level of a fermionic mode on $S^3$\,.

We also expect intuitively that across these transitions, the level
with  energy $\vep_\ell$ should become occupied. This can be
verified by computing the derivative of the action/free energy with
respect to $\mu$:
\be
\langle (N_L-N_R)\rangle 
\,= - T\,\frac{\partial S_{\rm eff}}{\partial \mu}\,=\, 
N_f\, N\, \sum_\ell d_\ell \, \zeta_\ell \, \int
d\theta'\,\rho(\theta')\,\frac{\zeta_\ell+\cos\theta'}{1+\zeta_\ell^2
  +2\zeta_\ell\,\cos\theta'} 
\ee
When $\mu < \vep_\ell$, $\zeta_\ell \ll 1$ at low $T$, and
therefore the corresponding energy level is unoccupied. For $\mu>
\vep_\ell$, on the other hand, since $\zeta_\ell\gg 1$, the
corresponding energy level has occupation number,
\be
N_f\, N\,d_\ell \, \zeta_\ell \, \int
d\theta'\,\rho(\theta')\,\frac{\zeta_\ell+\cos\theta'}{1+\zeta_\ell^2
  +2\zeta_\ell\,\cos\theta'} \,\approx \,N_f \,N \,d_\ell\,\qquad
\zeta_\ell \gg 1\,.
\ee

It is also interesting to note that 
in the ungapped phase, when $\rho (\theta) \approx
\frac{1}{2\pi}$, the Polyakov loop is
\be
\frac{1}{N}\langle{\rm Tr} U\rangle\,=\,
\int_{-\pi}^\pi\,d\theta\,e^{i\theta} \rho(\theta) \approx 0\,,\qquad
{\rm when}\,\, \zeta_\ell \gg 1\,\,{\rm or}\,\,\zeta_\ell \ll 1\,.
\ee
In the gapped phase, however, the Polyakov loop will have an
expectation value of order one. 
Therefore, in addition to a jump in the (chiral) 
fermion occupation number when $\mu_-^*\lesssim \mu \lesssim \mu_+^*$,
the Polyakov loop exhibits a spike across these transitons. The
situation is illustrated in Fig.\eqref{number}.

\Appendix{Finite density propagators}
\label{appb}
In this section we use real time fermion propagators at finite density
to sketch the perturbative computation of the correlator of the
operator  $\tilde\psi_q\psi_q$. We begin by quoting the results for
fermions in flat space \cite{boyanovsky}, and modify them slightly to
account for the fact that in our problem the chemical potential is for
an axial $U(1)$ symmetry. We use the flat space results 
for the two-point function of $\tilde\psi_q\psi_q$ to infer its form on 
$S^3$. We package the fermion flavours in the D3-D7 setup into $N_f$
Dirac fermions as in Eq.\eqref{dirac} $\left\{\psi^i\right\}$, 
$i=1,2,\ldots N_f$. The momentum space Feynman propagator is defined via
\be
i\,S(\vec k\,; t,0)\,\delta^l_m\,=\,
\int d^3x\, 
e^{-i\vec k\cdot \vec x}\,\langle \psi^l(\vec x,t)\,\bar\psi_m(0)\rangle\,.
\ee
where $l,m =1,2 \ldots N_f$. The Feynman propagator can be written as
the sum of retarded and advanced pieces
\be
S(\vec k; t, 0)\,=\,S^>(\vec k, t,0)\,\Theta(t)+S^<(\vec k; t,0)\,\Theta(-t)\,. 
\ee
For the case with an axial $U(1)$ chemical potential, we write
\be
S^{>(<)}(\vec k; t, 0) \,=\, \frac{1-\gamma_5}{2}\,S_-^{>(<)}(\vec k; t, 0)
\,+\,\frac{1+\gamma_5}{2}\,S_+^{>(<)}(\vec k; t, 0)\,.
\ee
With massless, chirality ``plus'' 
fermions, $k^\nu \equiv  (k, \vec k)$, we have 
\bea
S_+^>(\vec k ; t,0)\,&=&\,
-\frac{i}{2k}\,\left(\sh{k}\,[1-n_F(k,\mu)]\,e^{-i k t}
+\gamma_0 \sh{k}\gamma_0\,\bar{n}_F(k,\mu)\,e^{i k t}
\right)\\\nonumber
S_+^<(\vec k ; t,0)\,&=&\,
\frac{i}{2k}\,\left(\sh{k}\,n_F(k,\mu)\,e^{-i k t}
+\gamma_0 \sh{k}\gamma_0\,[1-\bar{n}_F(k,\mu)]\,e^{i k t}
\right)\,.
\label{finitemu}
\eea
The right-handed fermion propagators $S^>_-$ and $S^<_-$ 
are the same as above with
 the replacement $\mu\to -\mu$. The Fermi-Dirac distributions which
 turn into step functions at zero temperature are given as
\bea
&&n_F(k,\mu)\,=\, \frac{1}{e^{\beta(k-\mu)}+1}\, \to\, \Theta(\mu-k)\quad
{\rm as }\quad T\to 0\,,\\\nonumber
&&\bar n_F(k,\mu)\,=\, \frac{1}{e^{\beta(k+\mu)}+1}\, \to\, \Theta(-\mu-k)\quad
{\rm as }\quad T\to 0\,.
\eea
We are interested in computing the spatially homogeneous (zero
external momentum) part of the two-point function for ${\cal O}$,
\be
{\cal O} \equiv \sum_{i=1}^{N_f}\tilde\psi_{q\,i}\psi_q^i\,,\qquad
{\cal G}(t)\,\equiv\,\int d^3x\, \langle {\cal O}(\vec x, t)\,{\cal O}(0)\rangle\,.
\ee 
\\
\\
{\bf Time-ordered propagator:}
\\

It is straightforward to show that
\be
{\cal G}(t)\,=\,N_f\,\int \frac{d^3k}{(2\pi)^3}\,
{\rm Tr}[S(\vec k; 0,t)\,S(\vec k; t, 0)]
\ee
where the trace is over Dirac and colour indices, and all propagators
are time ordered. On $S^3$, the integral over
spatial momenta should be replaced by a discrete sum over the
spherical harmonic index $\ell$ with degeneracy $2\ell(\ell+1)$. The
Feynman propagator for ${\cal O}$ is then,
\be
{\cal G}(t)\,=\,N_f\int \frac{d^3k}{(2\pi)^3}\,
{\rm Tr}[S^>(\vec k; 0,t)\,S^<(\vec k; t,0)\, \Theta(-t)+
S^<(\vec k; 0,t)\,S^>(\vec k; t,0)\, \Theta(t)]\,.
\ee
After some algebra at $T=0$, we find, assuming that $\mu>0$, and the
number of colours $N$
\bea
{\cal G}(t)\,=\,N_f\,N\int\frac{d^3k}{(2\pi)^3}
&&\left(\left[ (1-\Theta(\mu-k))\,e^{2 i k t}\,+\,\Theta(\mu-k)\,e^{-2 i
      k t}\,\right]\,\Theta(-t)\right.\\\nonumber
&&\left.+\,\left[ \Theta(\mu-k)\,e^{2 i k t}\,+\,(1-\Theta(\mu-k))\,
e^{-2 i k t}\,\right]\,\Theta(t)\right)\,.
\eea
The frequency space propagator,
\be
\tilde{\cal G} (\omega) \,=\,i\,\int_{-\infty}^\infty dt e^{i\omega
  t}\,{\cal G} (t)\,.
\ee
is therefore,
\bea
&&\tilde{\cal G}(\omega)\,= N_f\,N\,\int \frac{d^3k}{(2\pi)^3}\,
\left[ \frac{1}{\omega+ 2k-i\epsilon} - \frac{1}{\omega-2k+i\epsilon}
\,+\right.\\\nonumber
&&\left.
+\Theta(\mu-k)\left(\frac{1}{\omega-2k-i\epsilon}
-\frac{1}{\omega+2k-i\epsilon}+\frac{1}{\omega-2k+i\epsilon}
-\frac{1}{\omega+2k+i\epsilon}
\right) 
\right]\,.
\label{polesum}
\eea
Rewriting the energy denominators using $1/(x-i\epsilon) = {\cal
  P}(1/x) + i\pi \delta(x)$, we get
\bea
\tilde{\cal G}(\omega)\,=\,N_f\,N\,\int 
\frac{d^3k}{(2\pi)^3}\,&&\left[\left({\cal
    P}\frac{1}{\omega+2k}
- {\cal P}\frac{1}{\omega-2k}\right)(1-2\Theta(\mu-k))\right.\nonumber\\\\\nonumber
&&\left.+ \frac{i\pi}{2} \delta(k+\frac{\omega}{2}) +\frac{i\pi}{2} \delta(k-\frac{\omega}{2})\right]\,.
\eea
Recalling that $k=|\vec k|$, the spatial momentum integrals can be
performed so that,  
\be
\tilde{\cal G}(\omega)\,=\,-\frac{N_f N}{16\pi^2}\left[\,\omega^2\,\ln\omega^2
+\,\omega^2\,\ln\left(\frac{\omega^2-4\mu^2}{\omega^2}\right)^2\right]\,.
\ee
Here we have ignored additive constants and contact terms, 
analytic in $\omega$.
The chemical potential has modified the $\mu =0$ result for the
Green's function of ${\cal O}$, introducing branch points at
$\omega=\pm 2\mu$.
\\
\\
{\bf Retarded propagator:}
\\

From the expression for the time ordered propagator, we see that the
contributions of on-shell intermediate states, even in the presence
of the chemical potential, only yield singularities on the real
$\omega$ axis. Therefore the different propagators - time-ordered and
retarded, differ only in the choice of $i\epsilon$ prescriptions
for these singularities. We confirm this by carrying out the explict
computation of the retarded propagator, defined as 
\be
\tilde{\cal G}_R(\omega)\,=\, i\int_{0}^\infty dt\,e^{i\omega
  t}\,\Theta(t)\,
\int d^3x\,\langle\left[{\cal O}(\vec x,t),{\cal O}(0)\right]\rangle\,.
\ee
We find that
\be
\tilde{\cal G}_R(\omega)\,=\,N_f\,N\int\frac{d^3k}{(2\pi)^3}\,
\left[\frac{1}{\omega+2k+i\epsilon}-\frac{1}{\omega-2k+i\epsilon}\right]
\left(1-2\Theta(\mu-k)\right)\,.
\ee
The real part of this expression (for real $\omega$) is exactly the
same as the time-ordered propagator. The main difference lies in the
imaginary part which is determined by the $i\epsilon$
prescriptions. Up to additive constants, the retarded propagator at finite
density in flat space is 
\bea
\tilde{\cal G}_R(\omega) = 
-\frac{N_f N}{16\pi^2}\left[\omega^2\ln\omega^2 
+\omega^2\ln\left(\frac{4\mu^2-\omega^2}{\omega^2}\right)^2
+2i\pi\omega^2{\rm sgn}(\omega)\Theta(\omega^2-4\mu^2)
\right]
\eea

We can also deduce the form of the two-point function of ${\cal O}$ on $S^3$
from the formula \eqref{polesum} by replacing the integrals by
a discrete sum over the energy levels on $S^3$:
\be
\tilde{\cal G}(\omega)= \frac{ N_f N}{2\pi^2R^3}\sum_{\ell=1}^\infty
\,d_\ell\,\left(\frac{1}{\omega+ 2 \vep_\ell}-\frac{1}{\omega-
2\vep_\ell}-
2\Theta(\mu-\vep_\ell)\left[\frac{1}{\omega+2\vep_\ell}
-\frac{1}{\omega-2\vep_\ell}\right]\right)
\ee
where for $S^3$ fermion harmonics
\be
\vep_\ell=(\ell+\tfrac{1}{2})\,R^{-1}\,,\qquad d_\ell = 2\ell(\ell+1)
\ee

\Appendix{D7-brane embedding ansatz}
In this appendix, we see why the embeddings with $g(z)\neq 0$ are not
suitable for the theory with massless hypermultiplets. With the
general ansatz
\be
\phi(z,t)\,=\,2\,\mu t + g(z)\,,
\ee
the DBI action for the probe D7-brane becomes
\bea
S_{\rm D7}\,=&&\, 2\pi^2 \,{\cal N}_{\rm D7}\,\int dt
\,dz\,\cos^3\theta\,
\frac{(1-\frac{z^2}{4})^3}{z^5}\,\times\\\nonumber\\\nonumber
&&\sqrt{(1+z^2\theta'(z)^2)
\left((1+\tfrac{z^2}{4})^2-4\mu^2\,z^2\,\sin^2\theta\right)+g'(z)^2\,
\sin^2\theta\,z^2\,(1+\tfrac{z^2}{4})^2}\,.
\eea
As usual, since the action does not depend on $g(z)$,
 we have
\be
\frac{\partial {\cal L}}{\partial g'} = 2\pi^2\,{\cal N}_{\rm D7}\,
c\, =\, {\rm constant}\,.
\ee
From this condition we find,
\be
g'(z)\,=\,c\, \sqrt{
\frac{(1+z^2\theta'(z)^2)
\left((1+\tfrac{z^2}{4})^2-4\mu^2\,z^2\,\sin^2\theta\right)}{\cos^6\theta\sin^2\theta(1+\tfrac{z^2}{4})^2(1-\tfrac{z^2}{4})^6\,z^{-8}-c^2}}\,
\frac{1}{\sin\theta(1+\tfrac{z^2}{4}) z}
\ee

The equation of motion for $\theta(z)$ is best obtained through the
Legendre transformed action
\be
S_{\rm LT}\,=\,S_{\rm D7}-2\pi^2\,{\cal N}_{\rm D7}\,\int dt\,dz \,
g'(z)\, c\,.
\ee
which simplifies to
\bea
&&S_{\rm LT}\,=\,2\pi^2\,{\cal N}_{\rm D7}\,\int
dt\,dz\,\times\frac{1}{z\sin\theta(1+\tfrac{z^2}{4})}\\\nonumber
\\\nonumber
&&\sqrt{
(1+z^2\theta'(z)^2)
\left((1+\tfrac{z^2}{4})^2-4\mu^2\,z^2\,\sin^2\theta\right)
(\cos^6\theta\sin^2\theta(1+\tfrac{z^2}{4})^2(1-\tfrac{z^2}{4})^6\,z^{-8}-c^2)}
\eea
From the equation of motion for $\theta(z)$, we find that the
asymptotic behaviour of the slipping angle near the boundary is 
\be
\theta(z)\,=\, \theta_{(0)}\, z + \theta_{(2)}\,z^3+\tfrac{1}{2}
\theta_{(0)}\,(1-4\mu^2)\,z^3 \, \ln z +\ldots,
\ee
so that the constant of the motion $c$ does not appear in the mass or
the condensate of the corresponding field theory operator. However, it
does make an appearance in $g(z)$ and therefore in $\phi(z,t)$,
which then has the asymptotic form,
\be
\phi(z,t) \to 2\mu t + \frac{1}{2} \frac{c}{\theta_{(0)}^2}\,z+\ldots
\ee 
For massless fields this implies a putative divergence in $\phi$, near
the boundary. This suggests that non-zero values of $c$ are
potentially problematic for configurations with massless
hypermultiplets and so $c$ must be set to zero for this case. (See
\cite{O'Bannon:2008bz} for a detailed explanation of related issues.)

\end{document}